\documentclass[a4paper,12pt]{article}
\pdfoutput=1
 \usepackage[latin1]{inputenc}
\usepackage[T1]{fontenc}
\usepackage{amsmath}
\usepackage{amsfonts}
\usepackage{amssymb}
\usepackage{color}
\usepackage{graphicx}
\usepackage{caption}
\usepackage{comment}
\usepackage{braket}
\usepackage{subcaption}%

 \newcommand{\be}{\begin{equation}}
	 \newcommand{\ee}{\end{equation}}
	 \newcommand{\ba}{\begin{eqnarray}}
		 \newcommand{\ea}{\end{eqnarray}}
		 
		   \newcommand{\bea}{\begin{eqnarray}}
			 \newcommand{\eea}{\end{eqnarray}}
 \newcommand{\nn}{\nonumber}

\begin{document} \title{Non-Abelian Vortices in Holographic Superconductors}

\author{Gianni Tallarita \\ \vspace{0.2 cm}\\
\normalsize \it  Departamento de Ciencias, Facultad de Artes Liberales, Universidad Adolfo Ibáñez,\\{\normalsize \it Santiago 7941169, Chile
 }
 \\
}

\date{\hfill}

\maketitle
\begin{abstract}
We find, by an appropriate extension of the standard holographic superconductor setup, static bulk solutions which describe holographic duals to non-Abelian vortices. In the core of these vortices a scalar field condenses, breaking a non-Abelian global symmetry which leads to additional zero modes called orientational moduli. These moduli appear in the bulk as Goldstone bosons associated to the condensation of a neutral scalar field.

\end{abstract}
\section{Introduction}
The gauge/gravity duality relates the large $N$ limit of strongly coupled gauge theories to classical gravitational theories in anti-de-Sitter \cite{Maldacena:1997re}. Therefore, untreatable strongly coupled problems can, in some cases, be mapped to solvable classical gravitational theories. This was applied in particular to high $T_c$ superconductors whose physics is mapped to that of a charged scalar field in the background of an anti-de-Sitter black hole \cite{Hartnoll:2008kx} \cite{Hartnoll:2009sz}. In these systems, gravitational solutions dual to vortices in the superconductors have been found, both in the probe and fully back-reacted limits \cite{Albash:2009iq}-\cite{Roychowdhury:2014gja}. These solutions, both for the superfluid and superconducting cases, are characterised by a flux tube which extends from an AdS horizon to its boundary, inside which a charged scalar field vanishes. The $3+1$ dimensional bulk flux tube solutions represent $2+1$ vortices in the dual field theory. The correspondence is easily seen geometrically as the dual vortices in the field theory are simply the boundary end-points of the full bulk flux tube solutions. The charge density and temperature of the bulk theory break the conformal symmetry of the holographic strongly coupled quantum theory and thus, within it, vortices are normalizable regular excitations.   Initially, such solutions were studied restricting to the probe limit. In this limit the charged matter sector along with the gauge fields do not back-react on the background geometry which is a fixed AdS Schwarschild black hole \footnote{Or an AdS-Reissner-Nordstrom black-hole depending on the setup.}. This approach fails in capturing the full physics at zero temperature in which gravitational effects cannot be ignored.  In \cite{Dias:2013bwa}, using a less conventional symmetry breaking mechanism, the full backreaction was included and important novel physics was unconvered. In particular the zero temperature limit was investigated and it was shown that the IR physics of these holographic vortices can be neatly captured by defect conformal field theory.  Numerous studies have extended the results of a single vortex to the case of more intricate spatial dependences. Of particular interest are the solutions corresponding to holographic vortex lattices \cite{Maeda:2009vf}, whose spatial dependence can be obtained analytically close to the critical magnetic field of the phase transition. A general lattice solution far from the critical point, valid at all temperatures including zero and taking into account the backreaction onto the gravitational sector is (to the extent of the author's knowledge) still an open problem (although see \cite{Adams:2012pj} for progress in this direction). \newline
	This paper studies a further extension of vortex holography devoted to non-Abelian vortices. Perhaps confusingly, these vortices are not related to non-Abelian extensions of the Abelian bulk gauge symmetry responsible for the superconducting phase transition. They are vortices (and more generally solitonic solutions, see for example \cite{Shifman:2015ama}) which possess orientational moduli on their world-sheets.  These moduli are caused by the condensation of an additional scalar field in the vortex core which breaks a non-Abelian global symmetry. Since their discovery \cite{AHDT} \cite{auzzi}, these vortices have been deeply studied (see for example \cite{S} - \cite{SA}). They are believed to be important in describing the confinement of electric charges in QCD via a system analogous (or dual) to the confinement of magnetic ``charges" in conventional superconducting models \cite{idea}. Whilst the idea is certainly appealing and, to some extent, supported by lattice evidence, the precise and realistic mechanism by which such flux tubes can form is still largely unknown and remains unproven. Large progress was made in \cite{Seiberg:1994rs} where supersymmetric Yang-Mills theories were shown to possess vacua in which magnetic monopoles can condense, but these models are still far from real world QCD.  The underlying common problem of models which study non-Abelian vortices, especially the more realistic non-supersymmetric ones, is that one is restricted from accessing the phenomenologically interesting strong coupling regime. In this way, this work provides a very direct route into the physics of non-Abelian vortices at strong coupling. This paper will make no contact with confinement, or make any QCD related phenomenological claim, it wishes to serve as a toy model for strongly coupled studies of non-Abelian vortices using holography. \newline
\indent In \cite{Dias:2013bwa} the backreaction of bulk vortices on the geometry was included and the dual theory was described by a defect conformal field theory (DCFT).  Upon including the backreaction in the setup of this paper we can therefore speculate that the dual theory will be captured by a DCFT in which the defect carries additional degrees of freedom. We will not investigate these interesting issues further in this paper and leave it as future work. \newline
The paper is setup as follows: section 2 introduces the system we wish to consider, the probe limit and the static background, section 3 presents the static solutions of this system, section 4 discusses its free energy paying attention in particular to how it compares with the standard vortex solutions found previously, finally in section 5 we find the solutions describing the orientational moduli of the dual vortices and provide our conclusions in section 6.


\section{The System}

	We will consider a system with $U(1)\times U(1)\times SU(2)$ gauge symmetry in the background of an anti-de-Sitter Schwarschild black-hole. A complex scalar $\psi$ of negative mass squared is coupled to one of the gauge symmetries (which we will think of as electromagnetism) thus realising in this sector the standard set-up for a holographic superconductor. A triplet of complex scalars $\chi^i$, also of negative mass squared, is coupled to the other $U(1)$ symmetry.  This triplet is also in the adjoint representation of the $SU(2)$ gauge symmetry. The additional $SU(2)$ gauge symmetry is what constitutes our ``spin" symmetry since it corresponds to a global $SU(2)$ symmetry in the dual theory. The presence of the additional $U(1)$ gauge symmetry is used to provide a chemical potential for the spin field in the dual theory. As presented the method used to do so may not be the simplest, after all a $U(1)$ gauge symmetry is already present and we could simply couple the spin field to it, or we could use the $t$ component of the $SU(2)$ gauge field. Why we chose not to pursue the latter of these two alternatives will soon be obvious to the reader: we will be considering the condensation of a neutral order parameter with respect to $SU(2)$, hence we require the gauge fields of this symmetry to vanish. With regards to the former case, we found that an additional $U(1)$ sector served to ensure firstly that the standard holographic superconductor setup would remain largely untouched (especially its holographic vortex solutions) and, secondly, that the presence of an additional tunable parameter would make it easier to find desired solutions numerically. Throughout the paper we work in the probe limit in which, to leading order, the background is fixed and suffers from no back-reaction from the gauge or matter fields.  \newline
	
	 It is well known that the holographic superconductor set-up does not require a symmetry breaking potential for a superconducting phase transition to occur. The coupling to the gauge field and the effects of the gravitational sector (even in the probe limit) are sufficient for a phase transition to occur in which, below a certain critical temperature, the black hole acquires scalar hair.  We want to keep the effects of a holographic superconductor in one of the $U(1)$ sectors and thus we will not add a symmetry breaking potential there. However, the two scalar fields are coupled by an interaction potential, inspired by Witten's superconducting string set-up \cite{Wsupcond}, which serves to ensure that upon condensation of the $\chi$ field, the gauged $SU(2)$ symmetry breaks to a $U(1)$ subgroup spontaneously. Without further ado, the action for this set-up is,

\be
S = \int d^4 x \sqrt{-g} \left(\mathcal{L}_{\psi}+\mathcal{L}_{\chi}+V(\psi,\chi)\right) + S_b,
\ee
where
\be
\mathcal{L}_{\psi} = -\frac14 F_{\mu\nu}F^{\mu\nu} -   (D_\mu[A]\psi)^*D_\mu[A]\psi- m^2_\psi|\psi|^2,
\ee
\be
\mathcal{L}_{\chi} =-\frac14 G_{\mu\nu}G^{\mu\nu}-\frac14 Tr\left( H_{\mu\nu}H^{\mu\nu}\right) -   (D_\mu[G,H]\chi^a)^*D_\mu[G,H]\chi^a- m^2_\chi|\chi^a|^2,
\ee 
$S_b$ are appropriate boundary terms to render the action finite, and
\be
V(\psi,\chi) = \gamma |\psi|^2 |\chi^a|^2+ \beta \left(|\chi^a|^2\right)^2.
\ee

The important probe limit is justified by considering a field redefinition of the form
	\be
	A_\mu \rightarrow  \frac{1}{e} A_\mu, \;\; \psi \rightarrow \frac{1}{e}\psi,\;\; G_\mu \rightarrow \frac{1}{g}G_\mu,  \chi^a\rightarrow \frac{\chi^a}{g}, \;\; H^a_\mu \rightarrow \frac{1}{g}H^a_\mu,
	\ee
and then taking $\gamma \rightarrow g\gamma$, $\beta \rightarrow g \beta$ and the limit $e=g \rightarrow \infty$, whereby the dynamics of gravity decouples from that of our gauge-matter. In the above we used the notation $|\chi^a|^2 =(\chi^a)^*\chi ^a$, and
\bea
&&F_{\mu\nu} = \partial_\mu A_\nu - \partial_\nu A_\mu, 
\\
&& D_\mu[A]\psi = (\partial_\mu -iA_\mu)\psi,\\
&&G_{\mu\nu} = \partial_\mu G_\nu - \partial_\nu G_\mu, 
\\
&& D_\mu[G,H]\chi^a = (\partial_\mu\chi^a -iG_\mu\chi^a+\epsilon^{abc}H_\mu^b\chi^c),\\
&& H_{\mu\nu} = \partial_\mu H_\nu - \partial_\nu H_\mu + \left[H_\mu, H_\nu\right],
\eea
which implies that all gauge couplings are set to one \footnote{With this choice of gauge couplings the gauge symmetry of the $\chi$ sector is enhanced to $U(2)$, however we keep the distinction apparent since we will be setting the $SU(2)$ gauge fields to zero later on.}. The dimensionless parameter $\gamma$ controls the coupling between the two scalar fields and a self-interaction term in the $\chi$ sector is added proportional to $\beta$, which has mass dimension zero. \newline

The background is that of an $AdS_4$ Schwarschild black hole with line element given by
\be\label{metric}
ds^2 =  \frac{L^2}{u^2}(-h(u) dt^2 +dr^2 + r^2d\theta^2) + \frac{L^2}{u^2 h(u)} du^2,
\ee
with $L$ the $AdS$ radius and 
\bea
&& h(u) = 1- u^3 \, , \;\;\; \;\;\; T =\frac{3}{4\pi L^2},
\eea
 with $T$ the black hole temperature. The dimensionless coordinate $u$ takes the range $(1 , 0)$, where $u=1$ describes the position of the black hole horizon and $u=0$ is the $AdS$ boundary. The coordinate $r$ with range on the positive semi-infinite interval describes the radial coordinate in the plane transverse to the AdS coordinate $u$, with $\theta$ the polar angle in this plane. 
  
  \subsection{Equations of motion}

Our goal is to be able to describe vortices in the dual theory supplemented by an additional $SU(2)$ neutral condensate appearing in the core of the vortex. For this purpose we will look for solutions setting the $SU(2)$ gauge fields to zero, $H_\mu^a=0$. We know from previous work on spatially dependent condensates \cite{Albash:2009iq} that both solutions in which the condensate is maximum at $r=0$, the so called droplet solutions, and the standard vortex solutions exist. The main idea is therefore to couple both kinds of solutions in a way in which they can co-exist such that the droplet of one sector sits in the core of the vortex of the other. Hence, using the following ansatz 
\bea
A_u = G_u= 0, \nn\\ 
A_r = G_ r = 0,\nn\\ 
A_\theta = r^2 a_\theta (u,r),\nn \\
G_\theta = g_\theta(u,r),\nn \\ 
A_t = a_0(u,r), \nn\\
G_t = g_0(u,r), \nn\\
 \psi = e^{i\,n\theta} \frac{r^n}{L^n} \rho (u, r),\nn\\
 \chi^a=e^{i\,k\theta} \chi(u,r)\delta^{a3},
\eea

where we have explicitly included the $r$ rescalings in the $\psi$ sector to avoid divergences at the origin, the equations of motion reduce to (from here on we set $L=1$ \footnote{We invite the reader to beware of the apparent dimensional incongruences this choice causes.})

\be\label{eomeq1}
\partial_u^2 \rho + \frac{1}{h(u)} \partial_r^2 \rho - \frac{2+u^3}{uh(u)}\partial_u \rho+\frac{1+2n}{rh(u)}\partial_r \rho +\left[\frac{a_0^2}{h(u)^2}+\frac{\left(-m^2_\psi +u^2 a_\theta\left(2n-r^2 a_\theta\right)+\gamma \chi^2\right)}{u^2 h(u)}\right]\rho=0
\ee

\be
\partial_u ^2 a_0 + \frac{1}{h(u)}\partial_r ^2 a_0 + \frac{1}{rh(u)}\partial_r a_0 - \frac{2r^{2n}\rho^2}{u^2 h(u)}a_0=0,
\ee

\be
\partial_u ^2 a_\theta + \frac{1}{h(u)}\partial_r ^2 a_\theta - \frac{3 u^2}{h(u)}\partial_u a_\theta+ \frac{3}{rh(u)}\partial_r a_\theta - \frac{2r^{2n}\rho^2}{u^2 h(u)}\left(\frac{n}{r^2}-a_\theta\right)=0,
\ee

\noindent in the $\psi$ sector, and

\be\label{chieom}
\partial_u^2 \chi + \frac{1}{h(u)} \partial_r^2 \chi - \frac{2+u^3}{uh(u)}\partial_u \chi+\frac{1+2n}{rh(u)}\partial_r \chi\nn
\ee
\be
+\left[\frac{g_0^2}{h(u)^2}+\frac{\left(-m^2_\chi r^2+ k^2u^2+u^2 g_\theta\left(-2k+g_\theta\right)-\gamma r^{2+2n}\rho^2\right)}{u^2 r^2 h(u)}\right]\chi + \frac{2\beta}{u^2 h(u)}\chi^3=0,
\ee

\be\label{g0eq}
\partial_u ^2 g_0 + \frac{1}{h(u)}\partial_r ^2 g_0 + \frac{1}{rh(u)}\partial_r g_0 - \frac{2\chi^2}{u^2 h(u)}g_0=0,
\ee

\be\label{gthetaeq}
\partial_u ^2 g_\theta + \frac{1}{h(u)}\partial_r ^2 g_\theta - \frac{3u^2}{h(u)}\partial_u g_\theta- \frac{1}{rh(u)}\partial_r g_\theta + \frac{2\chi^2}{u^2 h(u)}\left(k-g_\theta\right)=0,
\ee

in the $\chi$ sector. Note that the ansatz involves switching on a $\chi$ field pointing only in one direction (chosen to be the $3$-axis) of the internal space. This is the main reason why the resulting solution has orientational moduli, they related to the residual $U(1)$ invariance of rotations around this axis. The differences between the equations of motion of the scalar and gauge sectors, which are a-priori similar from the Lagrangian, can be attributed entirely to the chosen rescalings in the ansatz.

\subsection{Asymptotic behaviour and AdS/CFT dictionary}

Analyzing the scalar field equations asymptotically as $u\rightarrow 0$ we find a consistent behaviour of the form
\be
\rho \rightarrow \rho_1(r)u+\rho_2(r)u^{2}
\ee
\be\label{chiboundary}
\chi \rightarrow \chi_1(r)u+\chi_2(r)u^{2}
\ee
provided the standard choice \cite{Hartnoll:2008kx} for the scalar masses is made
\be
 m^2_\psi = m^2_\chi = -2.
\ee
 This choice is above the Breitenlohner-Freedman bound for both scalars. Given that both boundary expansion modes are normalizable we can choose what mode to use in order to describe the condensate in each sector. We choose to work with $\rho_1$ and set $\rho_2 =0$, similarly $\chi_2=0$ and we work with $\chi_1$. This ensures that the phase transition in which both scalars develop is spontaneous. \newline

Regarding the gauge fields we have asymptotically
\be
a_0 \rightarrow \mu(r) + u\rho^a(r)+\ldots,
\ee
\be
a_\theta \rightarrow a_\theta(r) + uJ_\theta(r)+u^2 \hat{J}_\theta+ \ldots,
\ee
and
\be
g_0 \rightarrow \mu^\chi(r) + u\rho^\chi(r)+\ldots,
\ee
\be
g_\theta \rightarrow g_\theta(r) + uJ^\chi_\theta(r)+\ldots,
\ee
with $\mu(r)$ and $\mu^\chi(r)$ the chemical potentials (the authors of \cite{Yokoi:2015qba} call $\mu^\chi$ the ``spin accumulation"), $\rho^a(r)$ and $\rho^\chi(r)$ the charge densities (our meaning of charge will be to associate $\rho^a$ to the usual electromagnetic charge, whilst $\rho^\chi$ is the charge associated to the additional $U(1)$), $a_\theta(r)$ and $g_\theta(r)$ related to the magnetic fields (again, the notion of ``magnetic" is merely a label for the second $U(1)$) and $J_\theta(r)$, $J_\theta^\chi(r)$ to the azimuthal currents of their corresponding $U(1)$ sectors. In particular, the actual magnetic field expression is $B=\frac{1}{r} \partial_r (r^2 a_\theta)$. The $SU(2)$ gauge symmetry in the bulk describes a global $SU(2)$ symmetry of the boundary theory. We engineer the system in the bulk to break this symmetry down to $U(1)$. Throughout the paper we work with spatially constant $\mu$ and $\mu^\chi$. This defines a scale-invariant temperature $\tilde{T}$ as $\tilde{T} = T/\mu$ so that the effective temperature changes are obtained at fixed $\mu^{\chi}$ varying $\mu$. \newline

\section{Solutions}\label{sols}

The normal phase solution is easily found to be 
\be
a_0 = \mu (1-u), \quad g_0 = \mu^\chi (1-u),
\ee
\be
a_\theta = g_\theta = \rho = \chi =0.
\ee
In this case, both $\mu$ and $\mu^\chi$ are not functions of $r$. However, we are interested in finding solutions in which both scalar fields condense, and for which the condensates have particular spatial dependences. In order to do so, we will set $g_\theta = k$ everywhere, which is a consistent solution of its equation of motion eq.(\ref{gthetaeq}). The important point is that this is an energetically finite value of $g_\theta$ since in the $\chi$ sector we do not need to regularize the $D_\theta$ covariant derivative at spatial infinity (as happens for the standard vortex) since it is the $\chi$ field that vanishes there. The choice implies that there is no ``magnetic" field in this sector.  \newline

Before solving the full two dimensional equations numerically we look at the behaviour of the scalar fields near the boundary, which will determine the behaviour of the holographic condensates. 

In the branch of solutions which we consider, for which $\mu^\chi \neq 0$ and
\be
\chi \rightarrow \chi_1(r)u+\chi_2(r)u^{2}
\ee
we have asymptotically
\be
\partial_r^2 \rho_1 + \frac{1+2n}{r}\partial_r \rho_1 + \left(\mu^2+2na_\theta-r^2a_\theta^2+\gamma\chi_1^2\right)\rho_1=0,
\ee
\be
\partial_r^2 \chi_1+ \frac{\partial_r\chi_1}{r}+\left((\mu^\chi)^2+2\beta\chi_1^2+\gamma(r^n\rho_1)^2\right)\chi_1=0,
\ee
\be
\partial_r^2 a_\theta +2\hat{J}_\theta+3\frac{\partial_r a_\theta}{r}+\frac{2}{r^2}\left(n-r^2 a_\theta\right)(r^n\rho_1)^2=0.
\ee
We recognise this set of equations as those obtained in \cite{S} (upon the appropriate field rescalings), which lead to the non-Abelian vortex solutions for the condensate profiles. Here we see why including a coupling to a $U(1)$ sector for the spin field is a necessity. If we set $\mu^\chi =0$ in the second equation we see that if $\beta\neq0$ then a constant core ($r=0$) value of $\chi_1$ is not allowed and if $\beta=0$ then the solution is $\chi\rightarrow C+\ln(r)+...$ which diverges logarithmically as $r\rightarrow0$. Therefore the chemical potential in this sector serves to stabilise the core value of the $\chi_1$ condensate. \newline


 Equations (\ref{eomeq1})-(\ref{g0eq}) are solved numerically using the COMSOL Multiphysics module. A far-radius cutoff $R = 20$ is used. We impose the following boundary conditions on the relaxation procedure
\be
a_0(1,r) = g_0(1,r) =0, 
\ee
\be
a_0(0,r) = \mu , \quad g_0(0,r) = \mu^\chi,
\ee
\be
\rho(0,r) = \chi(0,r) = 0,
\ee

and vanishing flux conditions at large and small $r$. This implies that we focus only on the branch of solutions with spontaneous symmetry breaking. \newline

\begin{figure}[ptb]
\begin{subfigure}{.5\textwidth}
\centering
\includegraphics[width=0.9\linewidth]{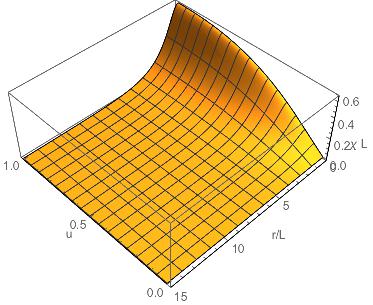}
\caption{$\chi(u,r)$}
\end{subfigure}
\begin{subfigure}{.5\textwidth}
\centering
\includegraphics[width=0.9\linewidth]{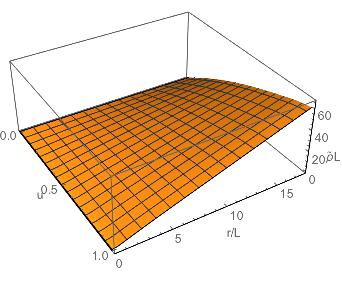}
\caption{$\rho(u,r)$}
\end{subfigure}
\begin{subfigure}{.5\textwidth}
\centering
\includegraphics[width=0.9\linewidth]{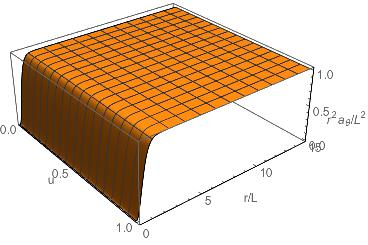}
\caption{$a_\theta(u,r)$}
\end{subfigure}
\begin{subfigure}{.5\textwidth}
\centering
\includegraphics[width=0.9\linewidth]{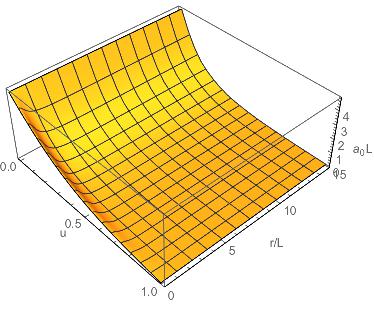}
\caption{$a_0(u,r)$}
\end{subfigure}
\begin{subfigure}{\textwidth}
\centering
\includegraphics[width=0.5\linewidth]{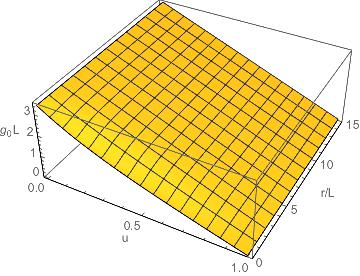}
\caption{$g_0(u,r)$}
\end{subfigure}
\caption{here we have $n=1$ for $\mu = 4.8$. $\tilde{\rho} = r^{2n} \rho / L^{2n}$ All at $\beta=-0.05$ and $\gamma=-0.23$, $\mu^\chi = 3.5$}%
\label{fig1}%
\end{figure}

\begin{figure}[ptb]
\begin{subfigure}{.5\textwidth}
\centering
\includegraphics[width=0.9\linewidth]{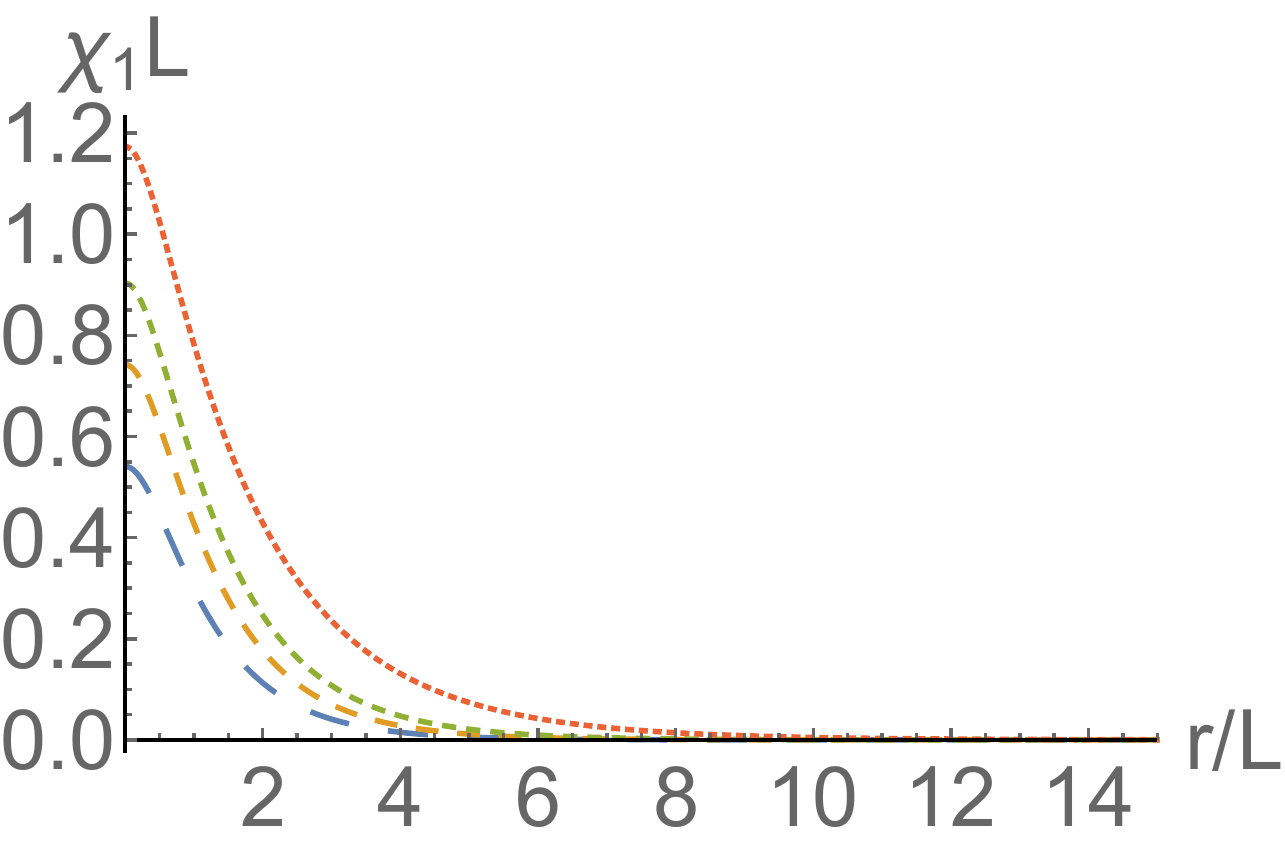}
\end{subfigure}
\begin{subfigure}{.5\textwidth}
\centering
\includegraphics[width=0.9\linewidth]{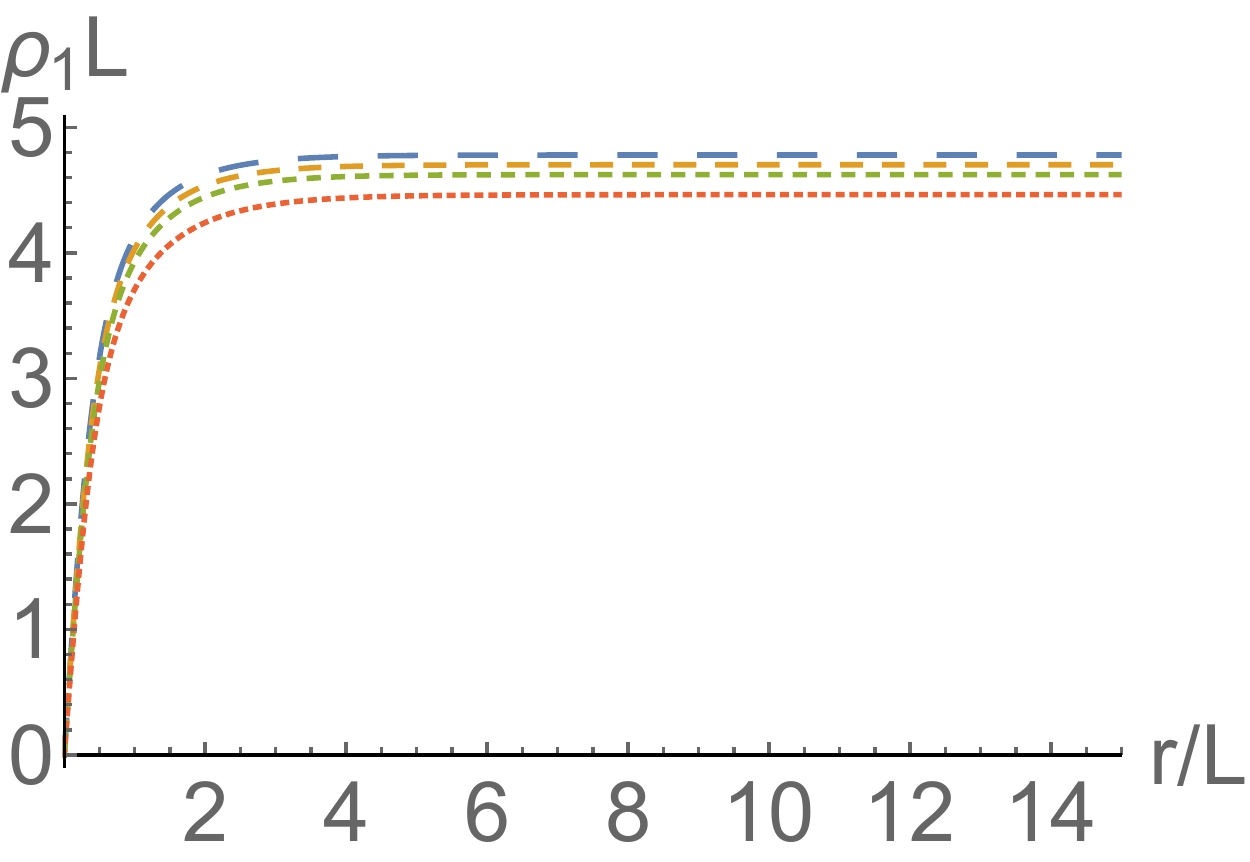}
\end{subfigure}
\begin{subfigure}{.5\textwidth}
\centering
\includegraphics[width=0.9\linewidth]{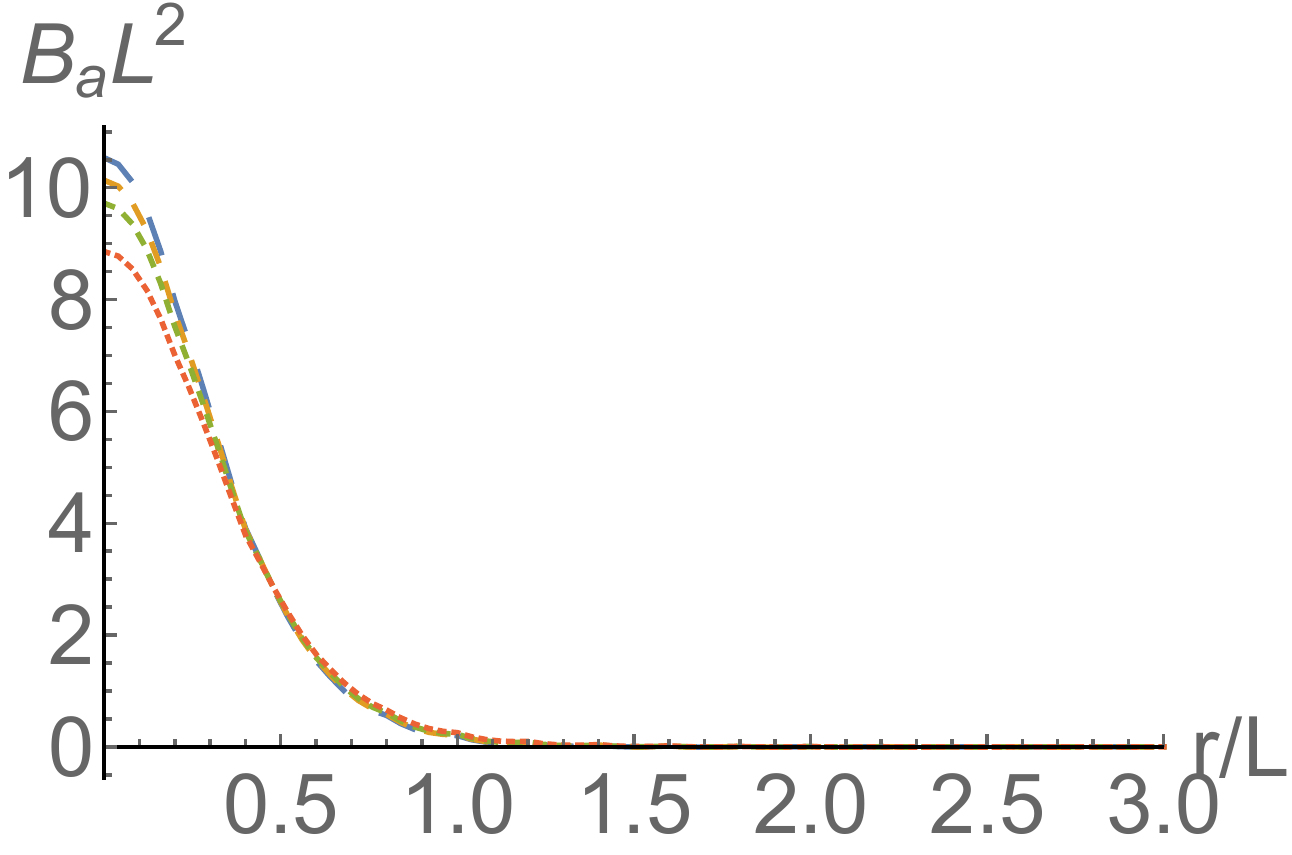}
\end{subfigure}
\begin{subfigure}{.5\textwidth}
\centering
\includegraphics[width=0.9\linewidth]{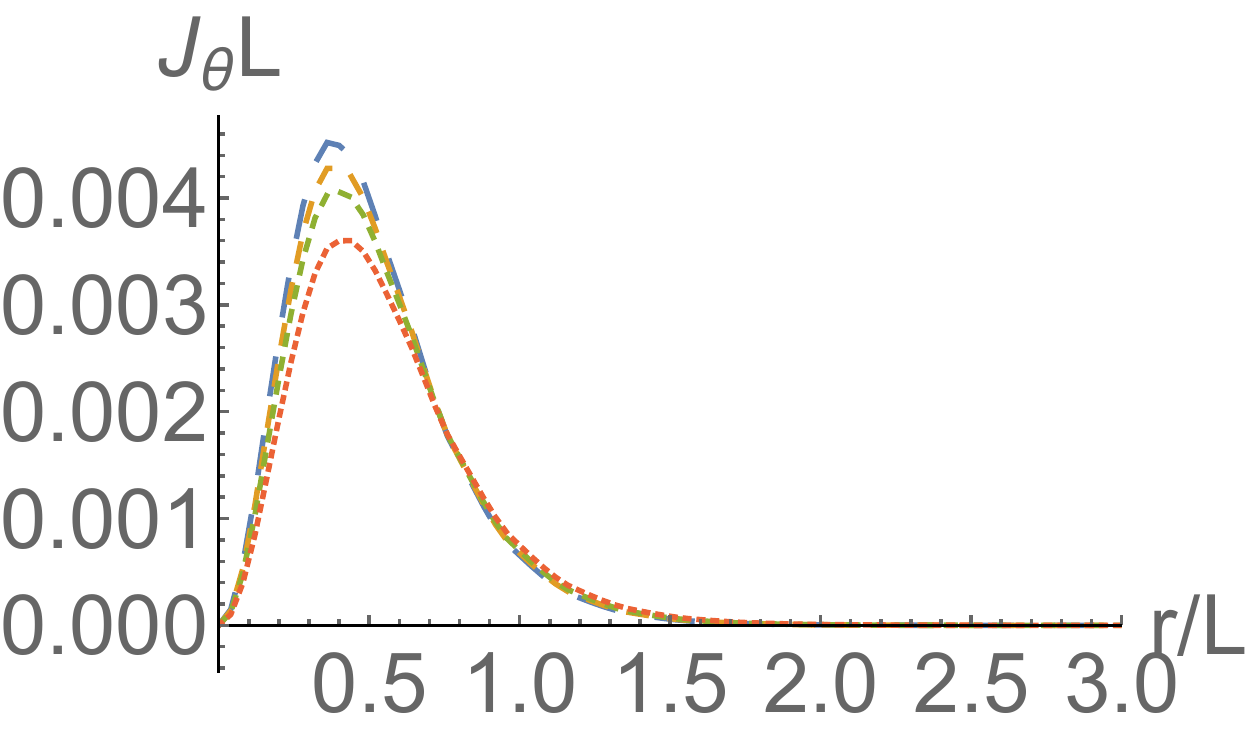}
\end{subfigure}
\begin{subfigure}{.5\textwidth}
\centering
\includegraphics[width=0.9\linewidth]{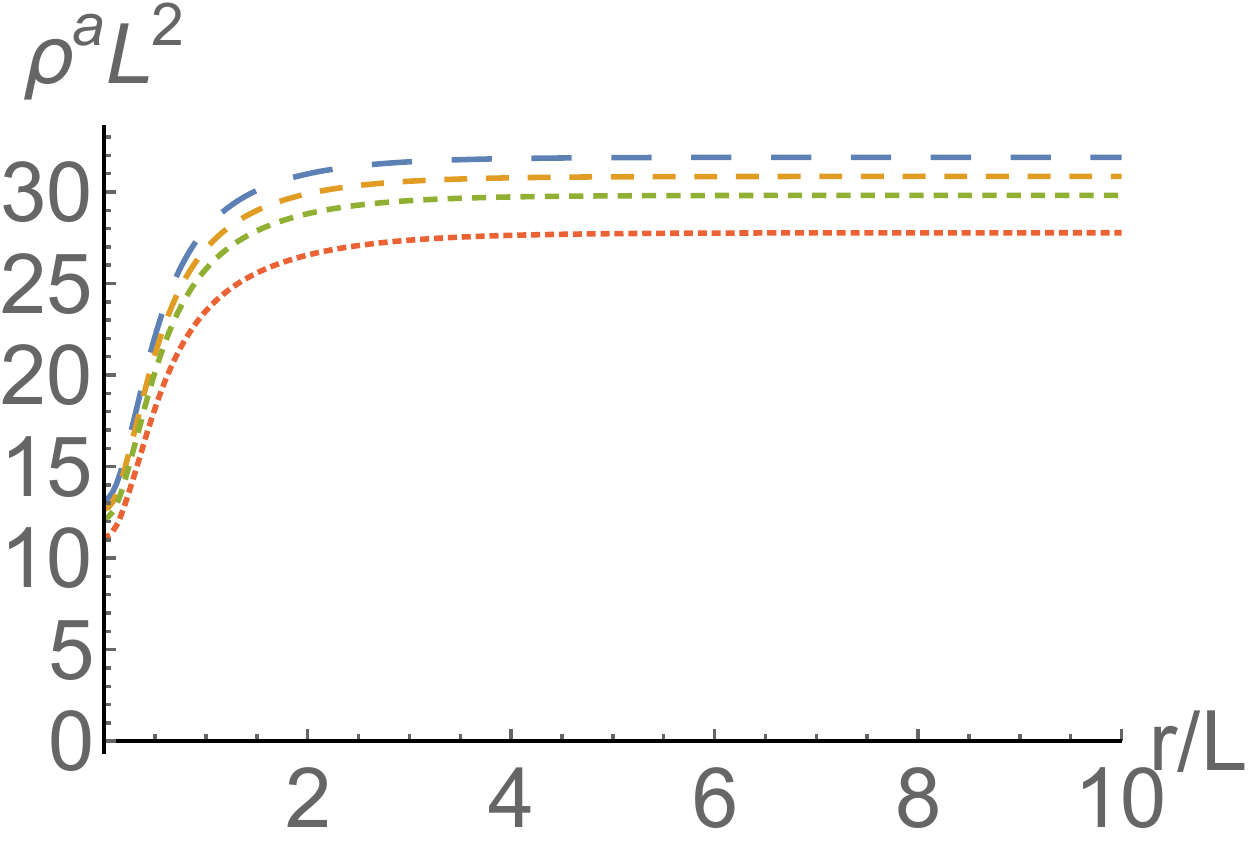}
\end{subfigure}
\begin{subfigure}{.5\textwidth}
\centering
\includegraphics[width=0.9\linewidth]{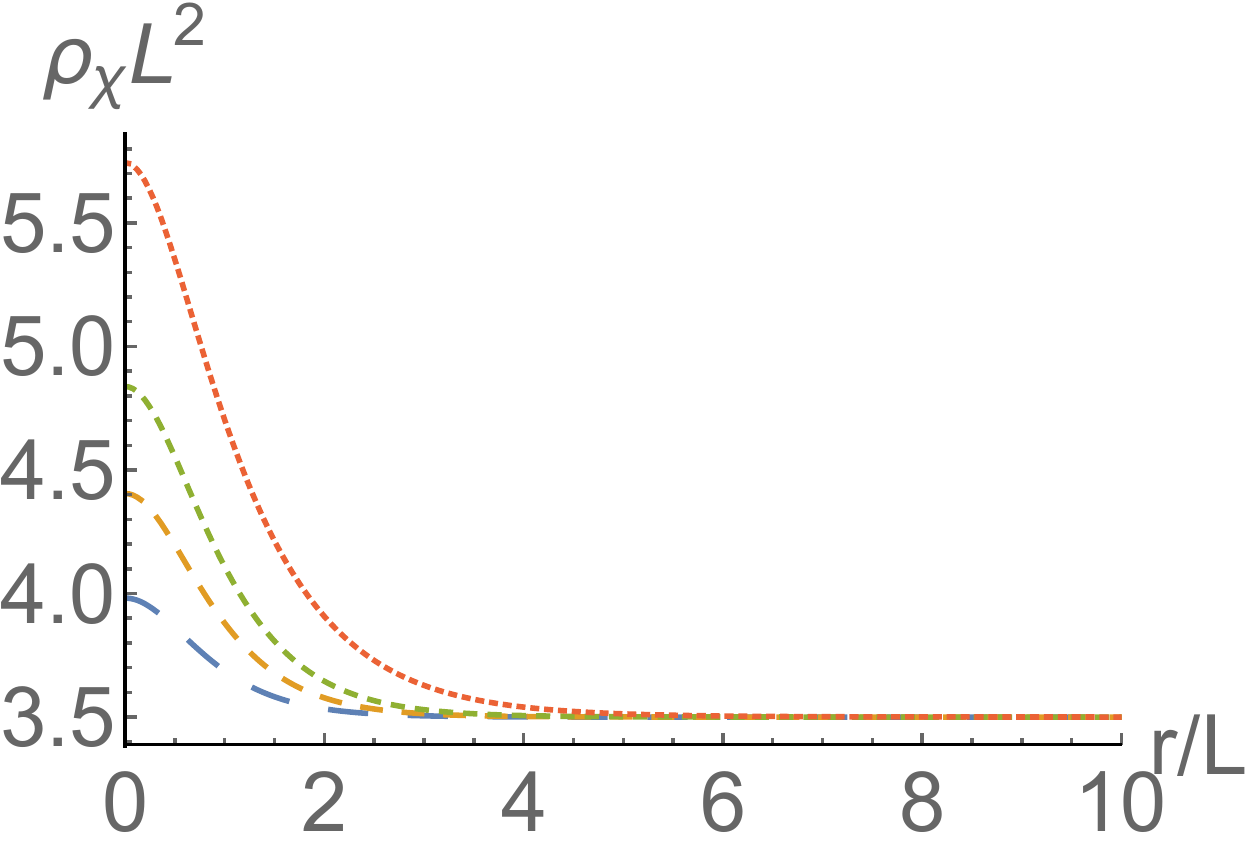}
\end{subfigure}
\caption{Field profiles at boundary for $n=1$ for $\mu = 4.8, 5.0,5.1,5.2$, red tiny dash is 4.8 blue large dash is 5.2. All at $\beta=-0.05$ and $\gamma=-0.23$, $\mu^\chi = 3.5$}%
\label{fig2}%
\end{figure}

\begin{figure}[ptb]
\begin{subfigure}{.5\textwidth}
\centering
\includegraphics[width=0.9\linewidth]{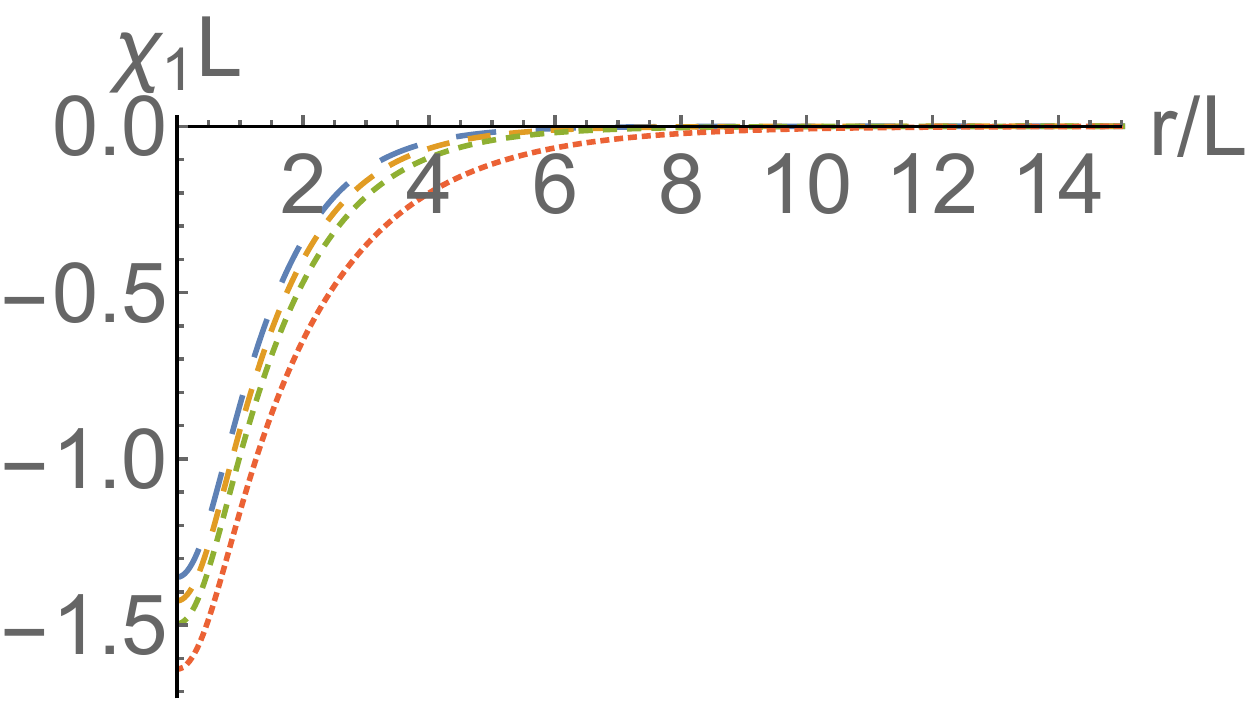}
\end{subfigure}
\begin{subfigure}{.5\textwidth}
\centering
\includegraphics[width=0.9\linewidth]{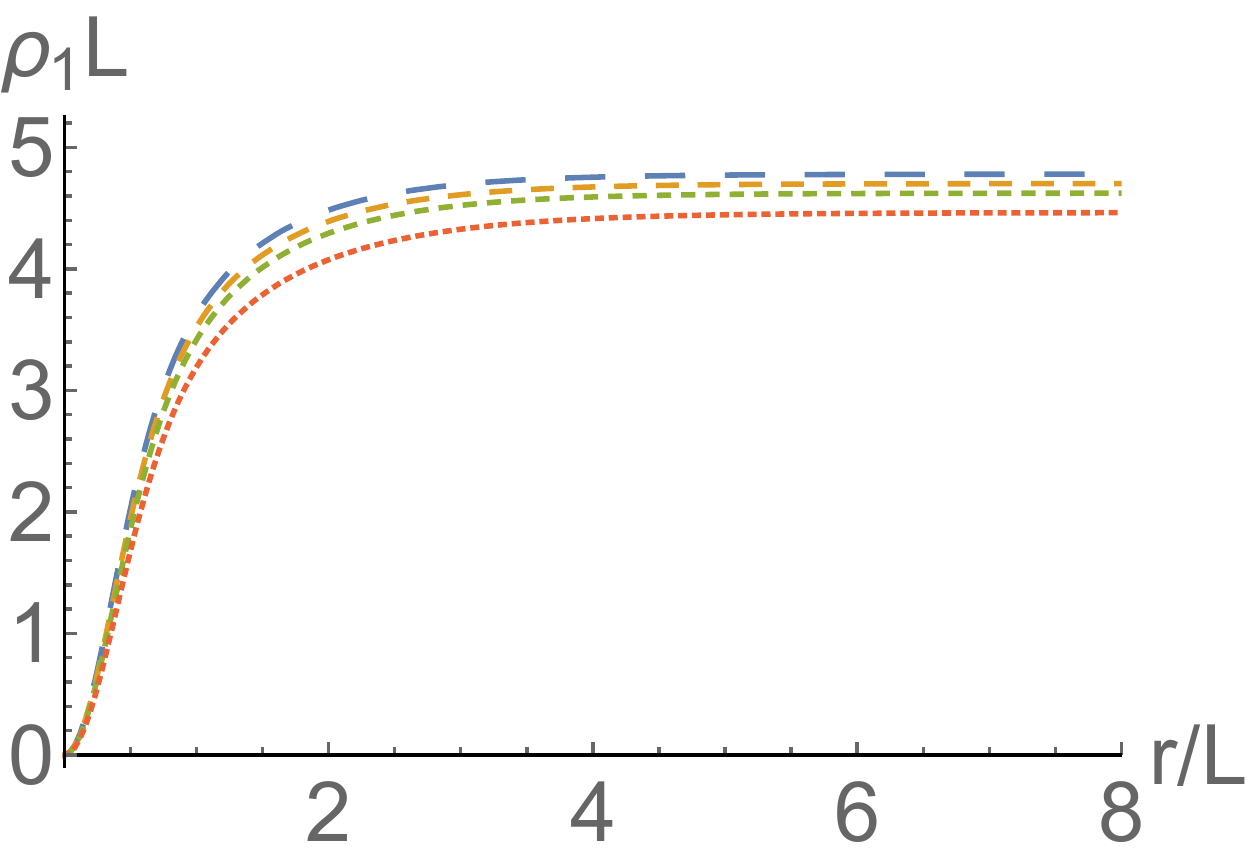}
\end{subfigure}
\begin{subfigure}{.5\textwidth}
\centering
\includegraphics[width=0.9\linewidth]{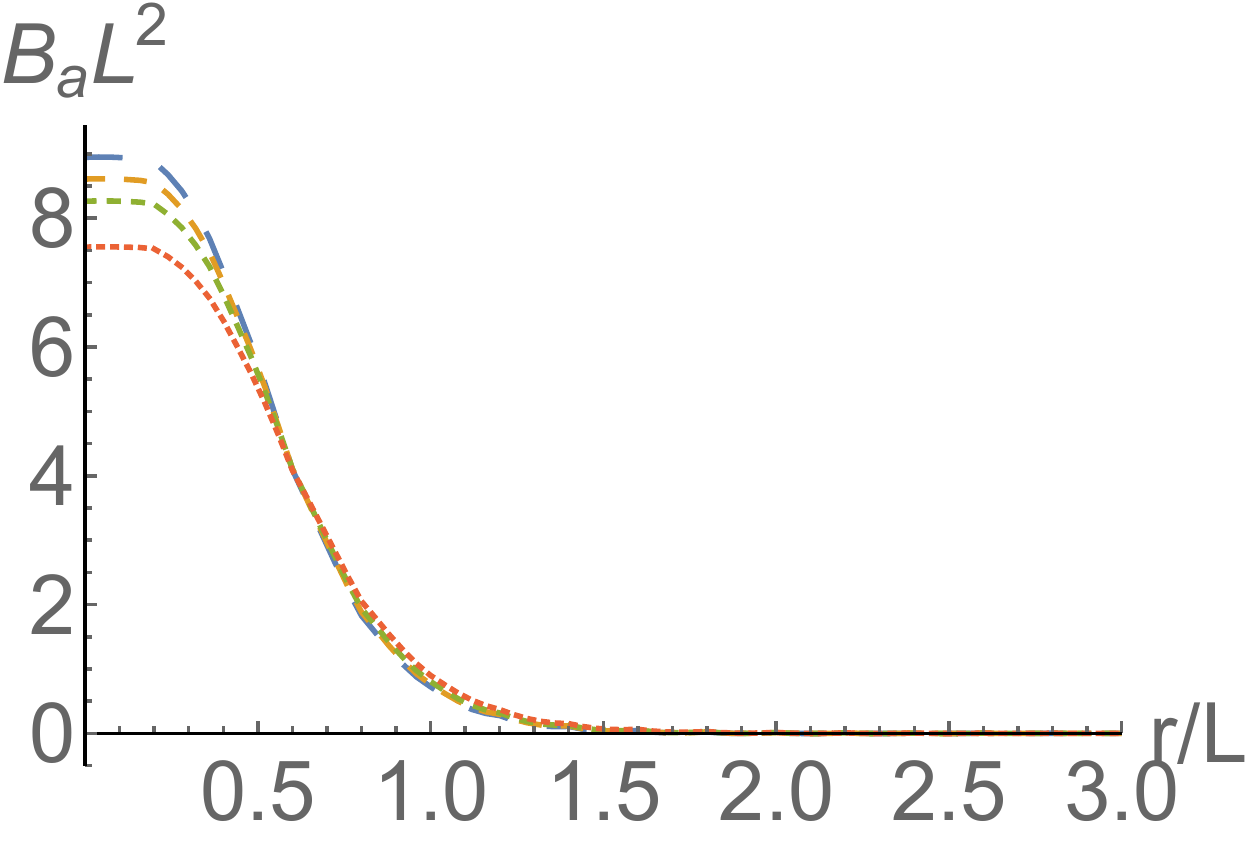}
\end{subfigure}
\begin{subfigure}{.5\textwidth}
\centering
\includegraphics[width=0.9\linewidth]{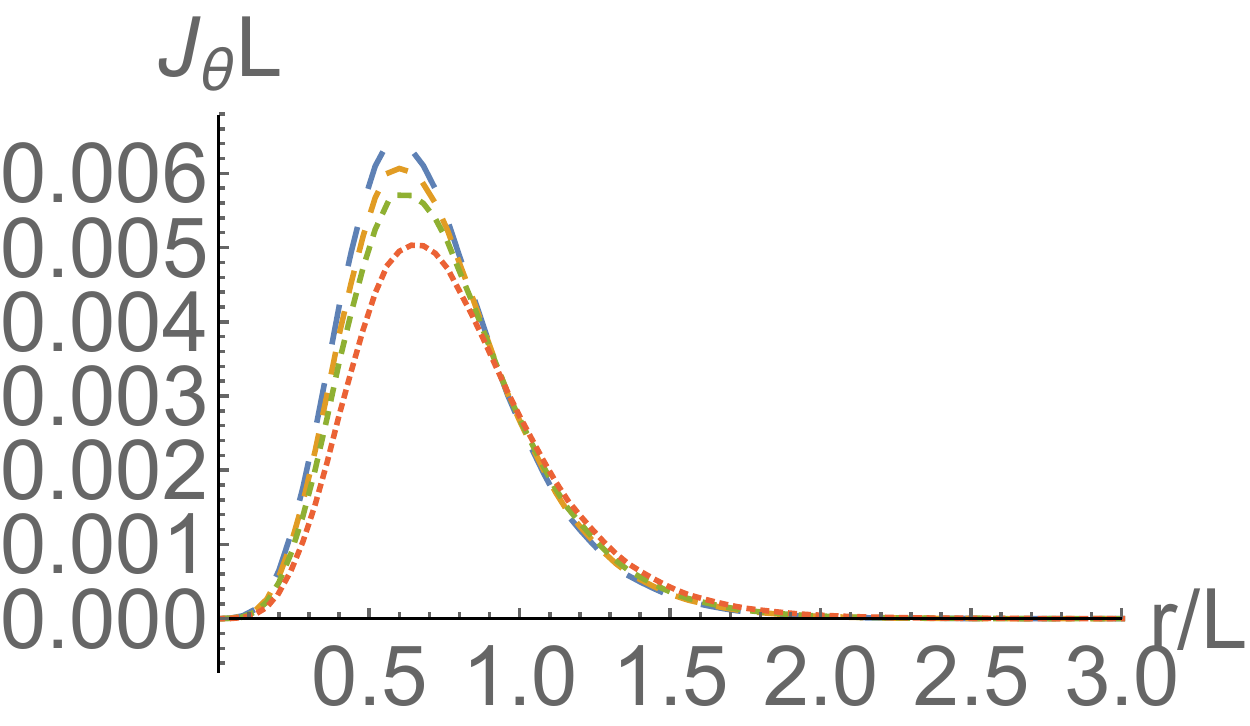}
\end{subfigure}
\begin{subfigure}{.5\textwidth}
\centering
\includegraphics[width=0.9\linewidth]{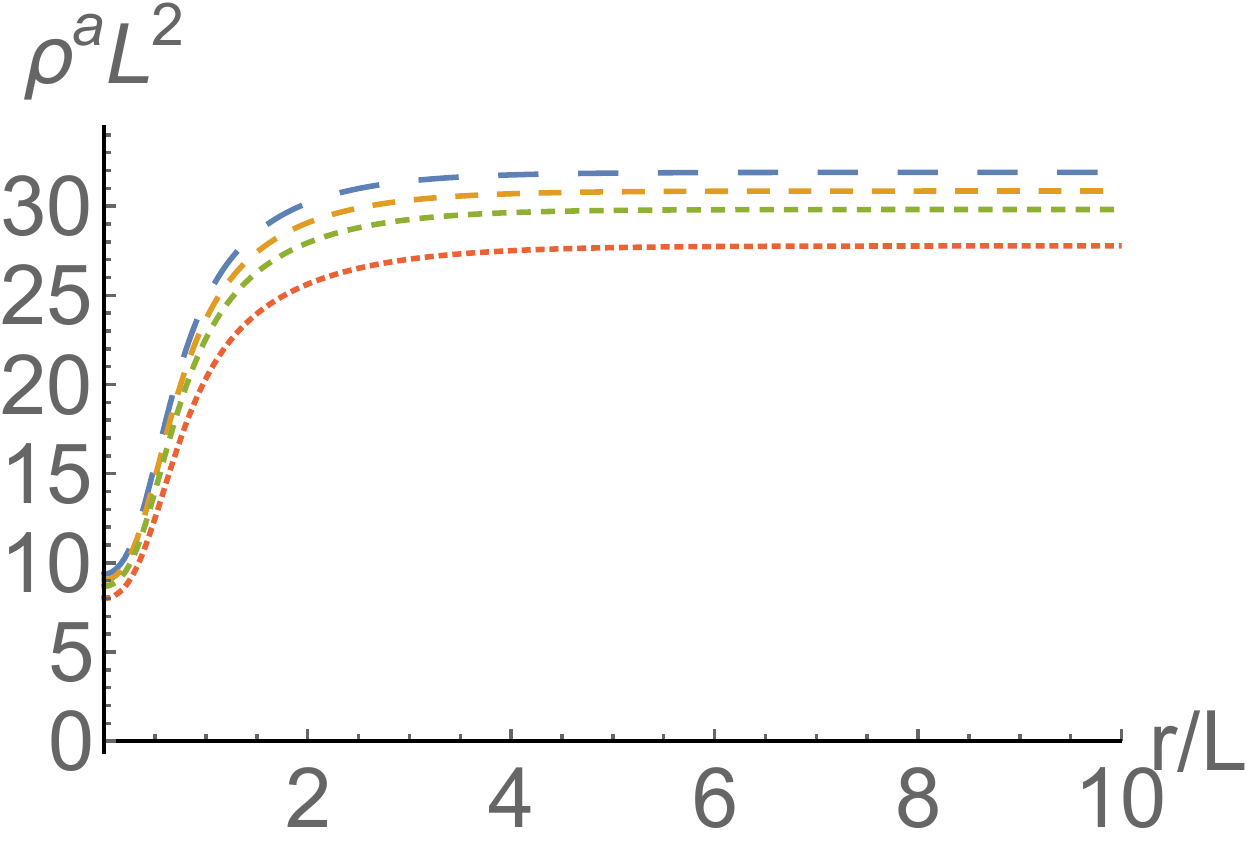}
\end{subfigure}
\begin{subfigure}{.5\textwidth}
\centering
\includegraphics[width=0.9\linewidth]{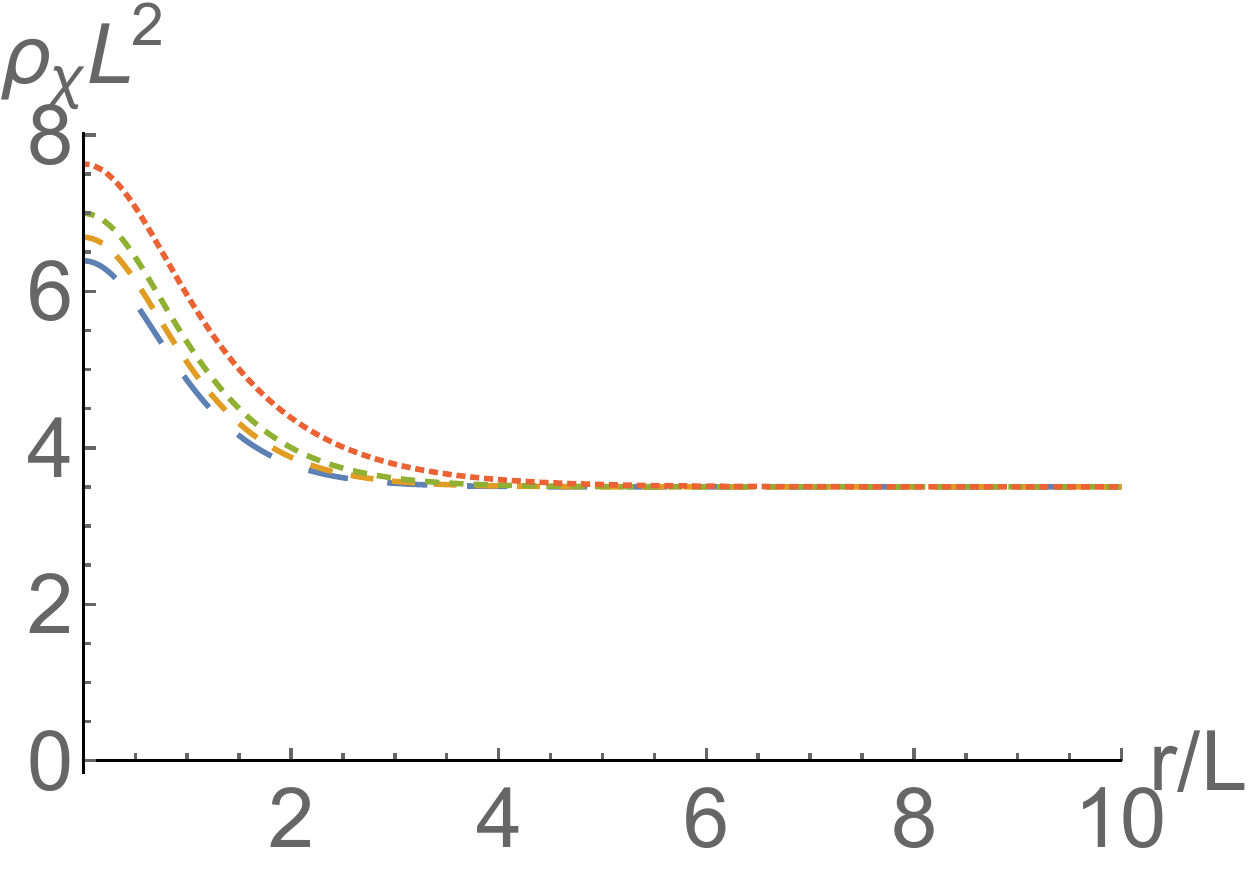}
\end{subfigure}
\caption{Field profiles at boundary for $n=2$ for $\mu = 4.8, 5.0,5.1,5.2$, red tiny dash is 4.8 blue large dash is 5.2. All at $\beta=-0.05$ and $\gamma=-0.23$, $\mu^\chi = 3.5$}%
\label{fig3}%
\end{figure}

The full two-dimensional solutions are shown in Figure \ref{fig1}. We present the relevant holographic quantities of interest in Figures \ref{fig2} and \ref{fig3} which show the $n=1$ and $n=2$ cases respectively (changing $k$ here does not affect the solutions as it is a trivial shift of the constant $g_\theta$ field). Note that the $a_\theta$ field has a small but non-vanishing dependence on the AdS coordinate $u$. We find in both cases that in the $\psi$ sector the condensate assumes a standard vortex form, whilst in the other the condensate is maximum in the core of the vortex. These solutions at the boundary resemble the flat space non-Abelian vortex solutions found in \cite{S}. The key point is the co-existence of both the ``droplet" (the $\chi$ sector) and ``vortex" ($\psi$ sector) solutions in standard holographic superconductivity.  \newline

As the temperature is raised (by lowering $\mu$) we find that the vortex condensate decreases in magnitude but the core condensate increases. Eventually, as the temperature is raised enough at around $\mu \approx 4.7$ we find solutions in which the $\chi$ condensate does not vanish at large $r$ and destabilizes the vortex solution in the other sector. In the opposite limit, where the temperature is made smaller, the $\psi$ condensate increases in magnitude and one must be careful in order to remain in the probe limit. For an accurate analysis of this phase at small temperatures one must include the effects of backreaction. 
 
In Figure \ref{fig4} we report on the variation of the $\chi_1$ field in the core as one varies the parameter $\beta$ in the potential. We find that increasing this parameter lowers the value of the condensate, as one generally expects from considerations in \cite{S}.

 \begin{figure}[ptb]
\centering
\includegraphics[width=0.6\linewidth]{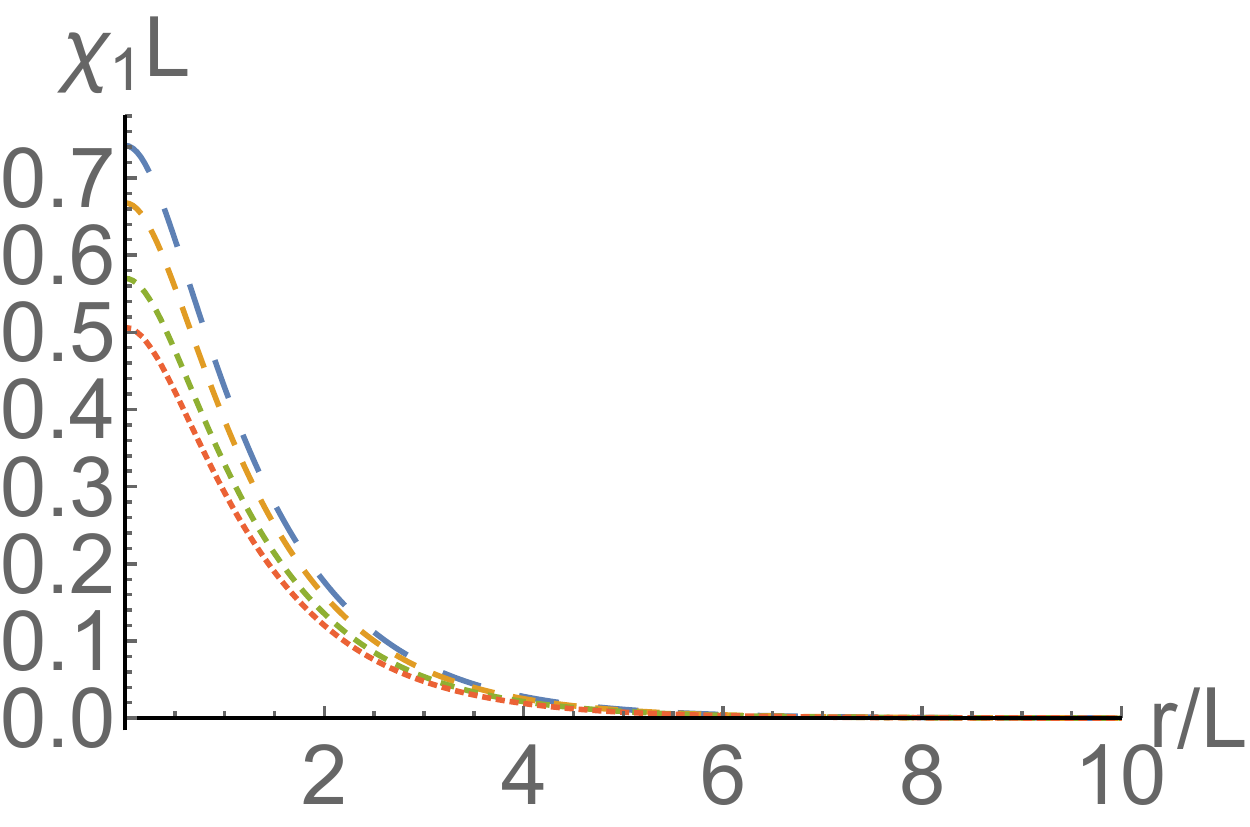}
\caption{Changes in $\chi_1$ for varying $\beta$. Plots are for $-\beta=0.05, 0.2, 0.5, 0.8$, with lowest $|\beta|$ corresponding to highest core value of $\chi_1$.}%
\label{fig4}%
\end{figure}

\section{Free energy}

In order to determine if this kind of condensed vortex phase is preferred over the normal un-condensed phase, we compute the difference of free energy densities (per surface area of boundary) between the condensed and normal phases by finding the on shell Euclidean action. The standard prescription relates the free energy $F$ to the Euclidean on-shell action $S^{os}_E$ as
\be
F = T S^{os}_E.
\ee

Since we are working with the operators dual to $\psi_1$ and $\chi_1$ we must add an appropriate boundary counter term to render the action finite, hence we set
\be
S_{b} = -\frac{1}{\pi R^2}\int d^2x \frac{h(u)}{u^2}(\psi \partial_z\psi+ \chi^i \partial_z\chi^i) \bigg|_{u=0}.
\ee

 The result of integrating by parts the scalar fields and using the equations of motion for the condensed phase is
 \be
 \frac{F_{cond}}{\pi R^2} = \frac{F_{bulk}+F_{surf}}{\pi R^2},
 \ee
where

\be
\frac{F_{bulk}}{\pi R^2} = \frac{2}{R^2}\int_0^R\int_1^0 \sqrt{g}\left[-\frac{1}{4}\left(F_{\mu\nu}F^{\mu\nu}+G_{\mu\nu}G^{\mu\nu}\right)-\left(\gamma|\psi|^2|\chi^i|^2+\beta\left(|\chi^i|^2\right)^2\right)\right]drdu,
\ee

\noindent and

\be
\frac{F_{surf}}{\pi R^2} = \frac{1}{\pi R}\int_1^0 \frac{1}{u^2}\left[r^{2n}\rho\partial_r\rho+n r^{2n-1}\rho^2 + \chi\partial_r \chi\right]\bigg|_{r=R} du.
\ee

The divergent contribution from the surface integral on the boundary of AdS space is cancelled by the counterterm. In the normal phase, where the solution is described by equations (\ref{eomeq1})-(\ref{g0eq}), the free energy density is simply

\be
\frac{F_{norm}}{\pi R^2} = -\frac{1}{2}\left(\mu^2+(\mu^\chi)^2\right).
\ee
Therefore, the difference in free energy densities is 
\be
\frac{\Delta F}{\pi R^2} = \frac{F_{bulk}+F_{surf}-F_{norm}}{\pi R^2},
\ee
so that, if the condensed phase is preferred, this quantity should be negative indicating that the condensed phase has lower energy (per unit area) than the normal phase. Note that for this comparison to make sense we must compare the phases at the same temperatures, i.e. both results should be compared at the same $\mu$ and $\mu^\chi$. The results of this comparison are shown in the table presented in figure \ref{fig5}.

\begin{figure}
\begin{center}
\begin{tabular}{|c|c|}
\hline
$\Delta \mathcal{F}/L^3$& $\mu L$\\
\hline
-29.5&5.2\\
\hline
-27.6&5.1\\
\hline
-25.7&5.0\\
\hline
-22.3&4.8\\
\hline
\end{tabular}
\end{center}
\caption{Free energy density difference between the condensed phase and the normal phase.}
\label{fig5}
\end{figure}

Numerically, the free energy difference is negative indicating that the condensed phase is preferred. As the temperature is raised this difference decreases, as one expects, tending to zero at the critical temperature. We tested the finiteness of these values for larger $R$ numerically and found stability of the reported values up to $R=25$ above which the numerical solver loses convergence. \newline

It is important to compare the free energy density of the solution with the extra spin field $\chi$ to the normal vortex solution without $\chi$. It was shown above that the condensed phase described by solutions with non-vanishing $\chi$ field is preferred over the normal phase, but it still might not be preferred over the normal vortex solution with no extra field. Hence, the appropriate quantity to compute is
\be
\frac{\Delta F}{\pi R^2} = \frac{F_\chi - F_\rho}{\pi R^2},
\ee
where $F_\chi$ denotes the solution with a non-vanishing $\chi$ field, and $F_\rho$ the standard vortex solution with no $\chi$. Since the $\chi$ field in the solution is small compared to the $\rho$ sector the normal vortex solution is only slightly altered by its presence for these values of $\mu$ and $\mu^\chi$ (clearly, if the temperature is lowered this is no longer true as the non-vanishing of the $\chi$ field at large $r$ has a significant effect on the vortex profile in the $\rho$ sector). Then we can safely approximate the contribution to the free energy of the $\rho$ sector to be equal in both the normal and non-abelian vortex phases (the validity of this approximation has been verified numerically). In this limit we have simply that
\be
\frac{F_{\chi}-F_\rho}{\pi R^2} = \frac{2}{R^2}\int_0^R\int_1^0 \sqrt{g}\left[-\frac{1}{4}G_{\mu\nu}G^{\mu\nu}-\left(\gamma|\psi|^2|\chi^i|^2+\beta\left(|\chi^i|^2\right)^2\right)\right]drdu
\ee
\be
-  \frac{1}{\pi R}\int_1^0 \frac{1}{u^2}\left[\chi\partial_r \chi\right]\bigg|_{r=R} du.
\ee
The numerical results of this computation are shown in Figure \ref{fig6}. The difference is once again negative, indicating that the solution supporting a non-vanishing condensate in the core is preferred to the normal vortex phase. In fact, this implies that the normal vortex solution in this system is at the most meta-stable. However, it does not imply complete stability of the solution with a $\chi$ field, only a stability analysis of the perturbation modes of this solution could determine this. We will not perform this analysis in this paper and content ourselves with the meta-stability of this solution implied by the energy arguments. A similar result was derived in \cite{Rogatko:2015awa} when considering vortices in the presence of an additional $U(1)$ gauge sector.

\begin{figure}
\begin{center}
\begin{tabular}{|c|c|}
\hline
$\Delta \mathcal{F}_\chi/L^3$& $\mu L$\\
\hline
-3.68&5.2\\
\hline
-3.28&5.1\\
\hline
-2.86&5.0\\
\hline
-1.89&4.8\\
\hline
\end{tabular}
\end{center}
\caption{Free energy density differences between the vortex with an additional $\chi$ condensate and the ``normal" vortex phase with $\chi=0$.}
\label{fig6}
\end{figure}

\section{Bulk localised Goldstone bosons as dual orientational moduli}

We begin the discussion regarding orientational moduli by reminding the reader of the well known case of Non-Abelian vortices in flat space (see \cite{SA} for an in depth explanation). In this case the field $\chi^i$ is a global triplet of $SU(2)$ and the pattern of symmetry breaking in the core of the vortex, where $\chi^3$ condenses, is given by $SU(2)\rightarrow U(1)$. The effective theory of the orientational moduli can be deduced topologically. It is given simply by
\be
\frac{SU(2)}{U(1)}\rightarrow CP(1),
\ee
which means that two orientational moduli, or Goldstone bosons of the global symmetry breaking, form a $CP(1)$ non-linear sigma model. These moduli live on the string world-sheet and therefore depend on the two coordinates $z$ and $t$, assuming the infinite string is aligned with the $z$ direction. The effective action governing these moduli can be easily obtained using the parametrization $\chi^i = \chi_0(r) S^i(t,z)$ with the condition that $S^i S^i=1$, and integrating over the background field $\chi_0(r)$ numerically. \newline 

The case considered here is more subtle. Recall that we are considering condensation of a scalar field which is originally charged under a gauged $SU(2)$ (we ignore the $U(1)$ charge in this discussion as it plays no role) and is made neutral by considering a specific solution in which the non-Abelian gauge bosons are set manually to zero. The condensation of this neutral scalar then causes the global symmetry breaking pattern $SU(2)\rightarrow U(1)$, the remaining symmetry related to rotations about the axis of the background field, and hence we expect topologically the same number of zero modes on the string and a similar $CP(1)$ non-linear sigma model. However to prove that these moduli exist we must show that we can find them as normalizable solutions in the bulk geometry which exist for $\omega \rightarrow 0$ in Fourier space, i.e. that they are gapless excitations. If these solutions exist, then they will be dual to Goldstone-modes in the field theory which, if localised on the vortex core, we will interpret as the orientational moduli of our vortex solution. In \cite{Iqbal:2010eh}, a very similar analysis was carried out in holographic models of antiferromagnetism to prove the existence of linearly dispersing spin waves. Even though the analysis here is similar, our intepretation will be different. We wish to find non-dispersing localised gapless modes corresponding to low frequency rotations of the order parameter in the core of the vortex. \newline

A-priori, we would expect the moduli to live on the world-line of our co-dimension 3 solitonic solution, and hence be fields which depend only on time $t$. 
To obtain the effective action a natural ansatz to consider is therefore, to a first approximation,
\be
\chi^i = \chi(r,u)S^i(t)
\ee
where $\chi(r,u)$ is the numerical solution obtained in the previous section and $S^i(t)$ is a time-dependent moduli field which satisfies $S^iS^i=1$. Inserting this into the action we obtain 
\be\label{act}
S_{0} = \int \sqrt{-g} g^{\mu\nu}\partial_\mu\chi^i\partial_\nu\chi^i \rightarrow I_1\int dt\; \dot{S^i}^2,
\ee
where
\be
I_1 =  2\pi\int drdu \frac{r}{u^2h(u)}\chi^2.
\ee
However, this integral is divergent at the horizon $u=1$ where $\chi$ condenses. This divergence indicates that the orientational moduli are not normalizable. Physically we could have anticipated it as this is just a consequence of the finiteness of the string in the $u$ direction. In the original variables related to $u$ by $z/z_h = 1/u$ the string is a semi-infinite object extending from the horizon $z=z_h$ to the boundary at infinity $z=\infty$. It indicates that we must consider also excitations of the moduli in the $u$-direction when calculating the effective action (see \cite{Cipriani:2012pa} for an explanation).  A natural choice would then be 
\be\label{rota}
\chi^i = \chi(r,u)S^i(t,u)
\ee
which leads to the effective action
\be\label{eff}
S = 2\pi\int dtdu \sqrt{-\tilde{g}}\tilde{g}^{\mu\nu} \left(\int dr\; r\chi(r,u)^2\right) \partial_\mu S_i \partial_\nu S_i.
\ee
with $\mu = (t,u)$ and $\tilde{g}_{\mu\nu} = (g_{tt}, g_{uu})$ from (\ref{metric}). Once again, the moduli fields are constrained by $S^i S^i = 1$. Clearly, it becomes complicated to observe the $CP(1)$ theory on the vortex world-sheet, in particular it is not straightforward to isolate the correct degrees of freedom corresponding to the moduli fields by integrating only in the $r$ direction. In particular, as noted in \cite{Iqbal:2010eh}, we must be careful in interpreting correctly rotations of the form (\ref{rota}). Expressions of this kind are space-time dependent rotations of the $\chi$ background field which in the context of the symmetries involved are simply gauge transformations of our adjoint vector. These gauge rotations must switch on vector field components as perturbations of our zero background solution which, with the parameterization chosen in eq.(\ref{rota}), enter as higher order corrections to the action. The complete action then becomes 
\be\label{fulllag}
\int d^4x \sqrt{-g} \left(\chi_0^2 D_\mu S^i D^\mu S^i -\frac{1}{4}H_{\mu\nu}^a H^{\mu\nu}_a\right),
\ee
where
\be
D_\mu S^a = \partial_\mu S^a +\epsilon^{abc} H_\mu^b S^c,
\ee
with $H_\mu^a$ a small perturbation over the vanishing background gauge field and $\chi_0 (u,r)$ our background solution. Determining the existence of the moduli fields then becomes a problem of solving the coupled linearized scalar-Yang-Mills equations in the gravitational background. The parameterization chosen here is not particularly suited for this problem as it involves solving the full non-Abelian equations. For this purpose, a better suited parameterization for the moduli fields exists, however this does not exclude that an appropriate solution can be found without switching parameterizations. In particular, inspired by \cite{Iqbal:2010eh} we set
\be\label{modpam}
\chi =\frac{1}{\sqrt{2}} \exp(i\pi^i(t,u)\tau^i)\chi_0(r,u),
\ee
where $\tau^i$ are the broken symmetry generators and $\chi_0(r,u) = \chi_0(r,u)\tau ^3$ is the background solution. The background solution is a function of the spatial variable aswell as the AdS coordinate $u$. Hence, the moduli fields are in this way automatically constrained to live in the vortex core (since $\chi_0(u,r)$ vanishes outside of the vortex core). This parameterization is appropriate in the sense that it allows us to find an explicit solution for the bulk Goldstone fields which survives as $\omega\rightarrow 0$ (as we will shortly show) however it is not well suited to discuss the non-linear sigma model these moduli belong to. The moduli fields here do not depend on the spatial variable $r$ and are in this way different from those proposed in \cite{Iqbal:2010eh}. The reason is that the gapless modes here are localised, whilst those in \cite{Iqbal:2010eh} defined a linearly dispersing spin wave. Importantly, this also means that we can ignore the perturbations in the spatial component of the gauge field, i.e. $H_r = 0$. Let us proceed to find these solutions, we follow the discussion of \cite{Iqbal:2010eh} closely. The quadratic action of the moduli fields is
\be
S_\pi=-2\pi\int dtdu \sqrt{-\tilde{g}}\tilde{g}^{\mu\nu}\frac{\tilde{\chi}_0^2}{2}\left(\partial_\mu \pi^i-H_\mu^i\right)\left(\partial_\nu \pi^i-H_\nu^i\right)+...
\ee
where $...$ denotes higher order corrections involving non-Abelian terms of the form $[\pi,H_\mu]$, and 
\be
\tilde{\chi}_0(u)^2 = \int dr r\chi_0(r,u)^2. 
\ee

The equations of motion which result from this action are (we drop the index $i$ on the moduli fields as they decouple, and the tilde on the metric components)
\be
\partial_\mu\left(\tilde{\chi}_0^2\sqrt{-g}g^{\mu\nu}\left(\partial_\nu\pi-H_\nu\right)\right)=0,
\ee
\be
\partial_m\left(\sqrt{-g} g^{\mu m}g^{\nu n}H_{\mu\nu}\right)+\sqrt{-g}g^{n\nu}\tilde{\chi}_0^2\left(\partial_\nu\pi-H_\nu\right)=0.
\ee

Note that a solution to these equations is the global rotation correpsonding to $\pi = \pi_0$ a constant and $H_\mu=0$. Let us begin by analysing the boundary behaviour of the scalar equation in the $\omega\rightarrow 0$ limit. We will work in Fourier space so that 
\be\label{pi1}
\pi(t,u)=\pi(u)e^{i\omega t}, \quad H_t(t,u) = H_t(u) e^{i\omega t},
\ee
and choose the gauge $H_u =0$. We know from equation (\ref{chiboundary}) that $\tilde{\chi}_0 \rightarrow Au + ...$ where $A$ is a constant. Then, if we ensure normalizability of the gauge sector by setting $H_\mu \rightarrow 0$ at the boundary we have that
\be\label{pi2}
\pi \rightarrow B + C u+...
\ee
with $B,\; C$ integration constants. From the expansion equation (\ref{modpam}) and the analysis of the $\chi$ boundary behaviour we see that both modes in the $\pi$ expansion are normalizable and we may interpret one as the source of the other. It is then a choice which source and vev to use. To work out the frequency dependent solution we expand about the constant global solution such that
\be
\pi = \pi_0 + \pi_1(u),... \quad \pi_1 \sim \mathcal{O}(\pi_0\omega^2),
\ee
\be
H_t(u)\sim \mathcal{O}(\pi_0\omega).
\ee
The infalling normalizable solution for $H_t$ is then 
\be
H_t = i\omega\pi_0\left(1-h_t(u)\right),
\ee
with $h_t$ the infalling solution of
\be\label{heq}
\partial_u\left(\sqrt{-g}g^{uu}g^{tt}\partial_u h_t\right)-\sqrt{-g}g^{tt}\tilde{\chi}_0^2h_t =0.
\ee
The normalizable condition on the gauge field requires $h_t(0)=1$. Then the solution for the moduli field reads
\be
\pi_1 = \omega^2\pi_0 f(u),
\ee
where $f(u)$ satisfies (from the $n=u$ component of the Yang-Mills equation)
\be
\tilde{\chi}_0^2\partial_u f = g^{tt}\partial_u h_t.
\ee
The scalar equation is then trivially satisfied\footnote{By ``trivially satisfied" we mean, in the notation of \cite{Iqbal:2010eh}, that $\alpha^t=0$.}. Therefore, a solution of equation (\ref{heq}) with $h_t(0)=1$ and $\partial_u h(1) =0$ describes a regular normalizable solution for the moduli fields. This solution, which exists in the $\omega\rightarrow 0$ limit, describes gapless modes related to low frequency rotations of our spin field, localised on the bulk flux tube, which we interpret as the orientational moduli of the dual theory vortex. \newline

A final comment on the gapless modes of our vortex needs to be made. We have so far considered the gapless modes corresponding to low frequency rotations of our ``spin" field in the vortex core and identified them as the orientational moduli of the dual non-Abelian vortex. However, the solution possesses one more gapless mode which we have so far ignored. This is just the mode corresponding to the un-eaten phase of the $\chi$ field in the standard holographic superfluid phase transition associated to the broken additional $U(1)$ gauge symmetry in the bulk. If we gave $g_\theta$ non-trivial dynamics on the boundary (as outlined in \cite{Domenech:2010nf}) and effectively gauged this symmetry the gapless mode would be eaten and our solution would have only the two orientational modes remaining. As it stands the dual vortex core describes a superfluid droplet, the superfluidity (or, once gauged, the superconductivity) being related to the additional $U(1)$ and not that of the original one which we called ``electromagnetism".

\subsection{The dual theory of the moduli}

Let us study the dynamics of the world-sheet theory a bit more and attempt to extract some precise information of the dual theory. Whilst the second moduli parametrization has revealed, in the vanishing $\omega$ limit, the existence of two gapless modes constrained in the vortex core it is the first of these parameterizations which is more suited to reveal some of the links to the dual theory.  The full dynamics of the scalar moduli world-sheet degrees of freedom, to linear order in the gauge field perturbations, is governed by the following equations
\be
\partial_\nu\left(\sqrt{-g}\tilde{\chi}_0^2g^{\mu\nu}D_\mu S^c\right)-\sqrt{-g}\tilde{\chi}_0^2g^{\mu\nu}\epsilon^{abc}\partial_\mu S^aH_\nu^b=0,
\ee
\be
\frac{1}{4}\partial_\mu\left(\sqrt{-g}g^{\mu\nu}g^{\tau\sigma}H_{\nu\sigma}^d\right)+\tilde{\chi}_0^2\sqrt{-g}g^{\mu\tau}\epsilon^{abc}\partial_\mu S^a S^c=0,
\ee
where $H_\mu$ is, as explained earlier, a small perturbation over the $H_\mu = 0$ background solution. In order to make some progress let us take the zeroth order approximation in which we can ignore the gauge field perturbations resulting in eq.(\ref{fulllag}) and work directly with eq.(\ref{eff}). As we will see, even this simplified system is not trivially resolved.  The dynamics of the moduli degrees of freedom of the world-sheet theory is determined by the equation of motion of the fields
\be
\partial_\mu\left(\sqrt{-\tilde{g}}\tilde{g}^{\mu\nu}\tilde{\chi}^2_0\partial_\nu S^i\right) =0,
\ee
subject to the additional constraint that $S^i S^i=1$. Once expanded out, the equation reduces to
\be
\partial_u \left(\frac{h(u)}{u^2}\tilde{\chi}_0^2\partial_u S^i\right)-\frac{1}{u^2h(u)}\tilde{\chi}^2_0\partial^2_tS^i=0.
\ee
This equation admits the separable solution $S^i (t,u) = f^i(u)g^i(t)$ (no summation intended), which gives the equations
\be
\frac{u^2h(u)}{\tilde{\chi}^2_0}\partial_u\left(\frac{h(u)}{u^2}\tilde{\chi}^2_0 \partial_u f^i\right)= A^i f^i,
\ee
\be
\ddot{g}^i =A^i g^i,
\ee
where $A^i$ is a constant vector. A full solution, to zeroth order approximation, is then to solve these two equations with ingoing boundary conditions simultaneously and imposing the constraint that $(g^if^i)^2=1$ coming from the identical constraint on $S^i$. This is a hard numerical problem as the function $\tilde{\chi}_0$ comes from the numerical solution of the background field equations for the vortex. In order to solve the above system we would therefore have to numerically integrate the solution in the $r$ direction and feed it as a seed to the $f$ equation which is then solved numerically with the additional constraint that $(g^if^i)^2=1$ must be satisfied everywhere. For now let us assume that a solution of this form exists (which after all was proven to exist at least in the low frequency limit by the other parameterization) and analyse the equation at the boundary, which is where we can make contact with the dual theory. Furthermore, analysing the solution at the boundary is representative of the full solution including the contribution from the gauge fields since they vanish there. In the $u \rightarrow 0$ limit, where $\tilde{\chi}_0 \rightarrow B u+...$ (see eq.(\ref{chiboundary})) where $B = \int dr r\chi_1(r)$, the equations reduce to the simple harmonic equations
\be
(f^i)''= A^i f^i, \quad  \ddot{g}^i =A^i g^i,
\ee
with the general solution
\bea\label{modb}
S ^i &=& \left(c_1 ^i \cos(a^i u)+c_2^i \sin(a^i u)\right) \times \left(c_3 ^i \cos(a^i t)+c_4^i \sin(a^i t) \right)\\
&=& \left(c_1^i+c_2^i a^i u +...\right) \left(c_3 ^i \cos(a^i t)+c_4^i \sin(a^i t) \right)
\eea
where $a_i = \sqrt{A}_i$, the $c_i ^j$'s are some integration constants and in the second line we have made manifest the $u\rightarrow 0$ limit. This solution is in some sense reassuring because we see that the moduli constraint can indeed be satisfied at least on the border, take for example the case in which $c^3_a =0$ then
\be
S^i S^i \approx (c_1^1c_3^1)^2 (\cos(a^1t))^2+(c_1^2c_4^2)^2 (\sin(a^2t))^2
\ee
where we set $c_4^1 = c_3^2 =0$. Clearly it is sufficient for the conditions $(c_1^1c_3^1)^2 = (c_1^2c_4^2)^2  = 1$ and $a_1 = a_2=\omega$ for $S^i S^i =1$ to hold in this limit, many other branches of solutions would equally satisfy the constraint, depending on the choices of the $c_i^j$'s. The classical energy of the solution, at least on the border, can be inferred from the original action in the $u \rightarrow 0$ limit. In general the energy density expression is simply
\bea\label{en2}
E &=& \frac{2\pi}{u^4} \tilde{\chi}_0^2\left(u^2 h(u) (g^i \partial_u f^i)^2+\frac{u^2}{h(u)}(\dot{g}^if^i)^2\right)\\
u\rightarrow 0&=& 2\pi B^2 \left((g^i \partial_u f^i\big|_{u=0})^2+(\dot{g}^if^i\big|_{u=0})^2\right).
\eea
By an appropriate choice of boundary conditions we can always pick $\partial_u f^i\big|_{u=0}=0$. Shortly we will discuss the holographic consequences of this choice. Then the energy density on the world-sheet theory in the $u\rightarrow 0$ limit reduces to 
\be\label{bounden}
E = 2\pi B^2 \left(\dot{g}^if^i\big|_{u=0}\right)^2 = 2\pi B^2 \left(c_1^i\dot{g}^i\right)^2.
\ee
We recognise this as simply the kinetic energy of the moduli fields, which is simply the rotational energy around the symmetry axis. There is a general moment of inertia vector which depends on the $c_1^i$, $I^i = 4\pi (\int dr r\chi_1) ^2 (c_1^i)^2$. This energy spectrum is continuous and labelled by the rotational frequencies in the moduli components. For the case above in which we considered $a_1 = a_2$ it can be further reduced to the single frequency $\omega$. Note that, in general, this energy needs to remain small in order for our probe approximation to remain valid which translates to considering small frequency rotations, i.e. $\omega << 1/L$ the only scale of the system. Luckily this is the right regime of validity of our previous solution and allows us to compare the two. Indeed the solutions obtained in both parameterizations should coincide at the border, and they do by a simple analysis of eq.(\ref{pi1}) and eq.(\ref{pi2}) translated to Fourier space \footnote{Working in Fourier space in the original $S^i$ parameterization is not simple since it would translate the constraint to a general convolution.}. The general energy expression derived in the other parameterization, evaluated on the solutions, leads to
\be
E = \frac{\pi}{2u^2}\tilde{\chi}_0^2\pi_0^2\omega^2\left(\frac{1}{h(u)}(h_t^2+\omega^4 f^2)+h(u)\omega^2 f'^2\right).
\ee
At the border, to leading order in $\omega$ we see that this expression reduces to
\be
E = \frac{\pi}{2}B^2\pi_0^2\omega^2+...
\ee
which confirms our analysis leading to eq.(\ref{bounden}) stating that the energy at the border is just the contribution from the rotational degrees of freedom. In the bulk we must also consider the contribution coming from the $\partial_u f$ terms, which is simply the momentum carried in the longitudinal modes along the string. This becomes important when discussion dissipation below. Indeed picking $\partial_u f=0$ at the border is a simple restatement that the longitudinal momentum is purely in-falling. \newline

Now let us try and learn what we can of the dual theory. The moduli fields $S^i(t,u)$ are holographically dual to operators $\mathcal{O}^i(t)$ responsible for gapless excitations on the dual vortex world-line. Through the boundary analysis carried out above we can see that both the boundary expansion modes of the $S^i$ field are normalizable (in much the same way as those of the $\pi^i$ field) and hence we may choose which one to interpret as the source and which as the vev. Let us say we pick the $c_1^1$ term as the vev and set $c_2 =0$, which is the choice corresponding to the above energy consideration. Then we can immediately read off from eq.(\ref{modb}) that  $\braket{\mathcal{O}^i(t)} = c_1^1 \left(c_3 ^i \cos(a^i t)+c_4^i \sin(a^i t) \right)$. We can interpret this as the operator $\mathcal{O}$ creating (gapless) excitations with energy $\omega$. Note in particular that using the same constraints on the $c_i^j$'s as derived above we also have that $\braket{\mathcal{O}^i(t)}\braket{\mathcal{O}^i(t)}=1$, and so these operators obey a similar constraint to the bulk field $S^i$ to which they are dual. This should be unsurprising since the presence of a $CP(1)$ non linear sigma model on the dual vortex world-line is a necessary result of the symmetry breaking pattern. It is just the result of the $SU(2)\rightarrow U(1)$ global symmetry breaking of the dual theory. It is however a nice result that we may see it directly from boundary considerations of our bulk world-sheet action. \footnote{In fact it is tempting to go one step further, since there is only one viable Lagrangian description of the one $0+1$ dimensional $CP(1)$ effective theory in flat space we can speculate that the dual action should be simply $S_{dual}\propto \int dt \dot{\braket{\mathcal{O}}}^i \dot{\braket{\mathcal{O}}}^i$, with the previous constraint imposed. Even though the original system is not directly derived from any string  framework for which a dual Lagrangian description is known, there seems to be sufficient symmetry to constrain the effective theory of the vortex enough for its Lagrangian to be inferred.}\newline

 \begin{figure}[ptb]
\centering
\includegraphics[width=0.5\linewidth]{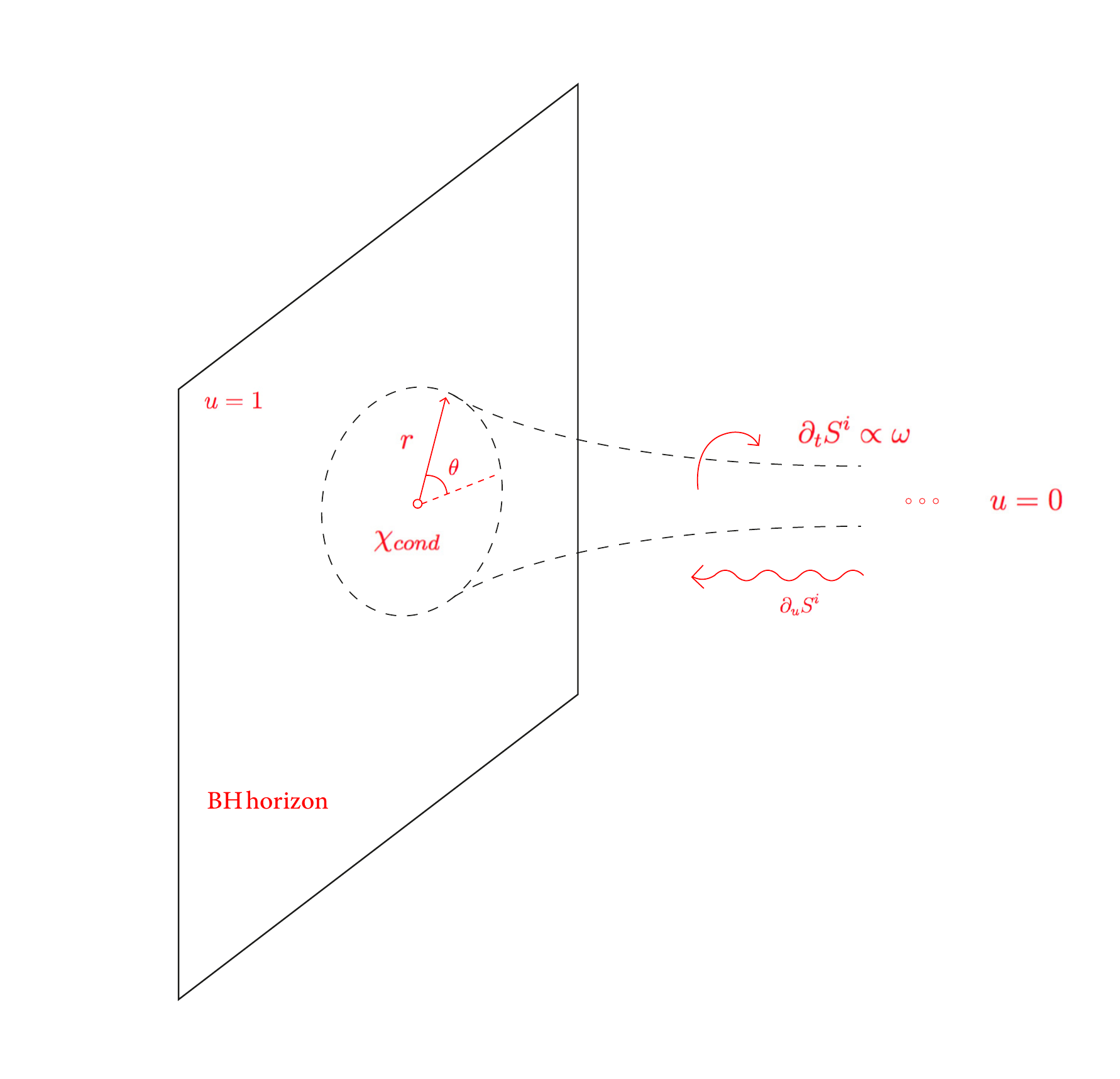}
\caption{Schematic diagram of vortex configuration with core excitations. Those proportional to $\omega$ involve the rotor degrees of freedom whilst the in-going $\partial_u S^i$ excitations are longitudinal momentum modes. These are infalling and meet the $\chi$ condensate at the black-hole horizon.}%
\label{fig4}%
\end{figure}

 \begin{figure}[ptb]
\centering
\includegraphics[width=0.5\linewidth]{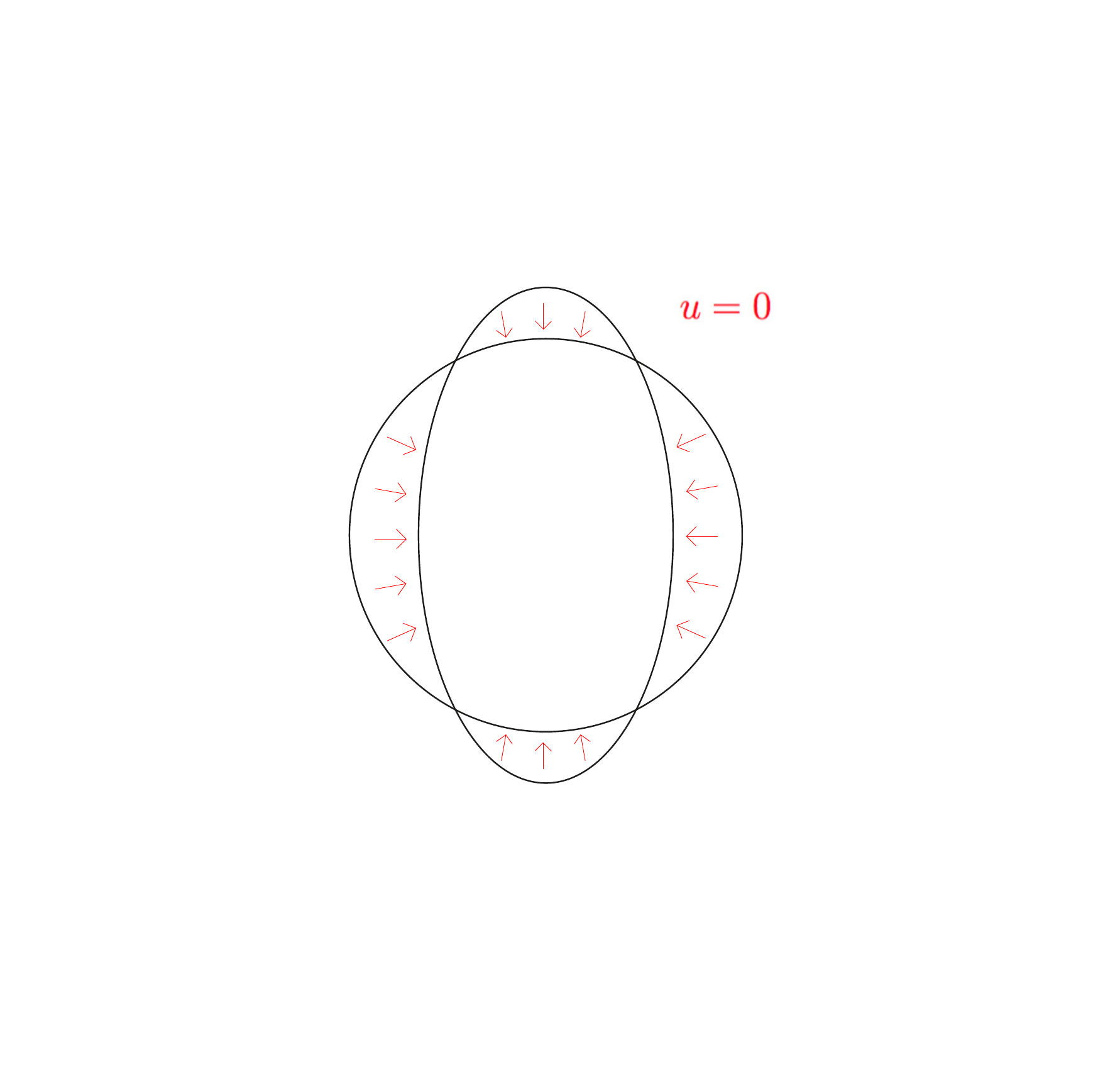}
\caption{Changes in shape of the vortex core, as a result of momentum transfer from the longitudinal modes of the vortex to the condensate at the horizon.}%
\label{fig4}%
\end{figure}

According to this picture the dual theory of the moduli degrees of freedom corresponds to a quantum rigid rotor system in a plane. As is well known the system has quantized energy levels
\be
E_s = \frac{1}{2I_d}\left(s(s+1)\right),
\ee 

with $s$ an integer and $I_d$ the moment of inertia of the dual rotor (in the isospace). To make sense of this picture recall that the probe limit forces us to consider only the small $\omega$ limit. In this limit the excitations have small energies compared to the gapped spectrum and therefore insufficient to probe the discrete energy levels. We can interpret these gapless excitations as small energetic perturbations from the ground state. Upon including the backreaction in the system, where we are allowed to probe all scales of $\omega$ then we can in principle consider excitations which are energetic enough to excite the system to its first energy level. This happens at $\omega \approx 1/I_d$.  We do not have a direct way of obtaining $I_d$ from the bulk data. It is tempting to conjecture that $I_d$ should be equal to $|I^i|$, the classical bulk moment of inertia evaluated on the boundary. This allows to get a numerical estimate by picking values of $c_1^i$.  Our solution at $L=1$ has $B \approx 1$, where $B$ was defined above. This is of course a quantity that depends on temperature and the various parameters of our system, but for the temperature ranges investigated in the paper it is always a quantity of order 1. Then, already for $c_1^1=c_1^2 = 0.5$,  $1/I_d \approx 1/\pi \approx 0.32$ which is comparable to $L$. Since the probe limit imposes $\omega << 1/L = 1$ we are far below the excitation energy. \newline

The last ingredient for this picture to make sense is to consider the dissipation of these excitations. There should be some mechanism by which energy is lost, otherwise the system would have a continuous spectrum of excitations. In the bulk the dissipation picture is easily understood. As argumented before, the finiteness of the string implies that we cannot ignore the momentum modes along the string direction. Then the moduli rotations can dissipate energy by exciting these low energetic longitudinal modes. These momentum modes are characterized by the $u$ dependence of $S^i$ and show up in the energy functional eq.(\ref{en2}) when not considered on the border. The in-falling boundary conditions then state that the momentum carried by these modes is invariably absorbed at the black hole horizon. At the horizon, in the core of the vortex, we have the condensate of the scalar $\chi$ field. If the energy reaching there was sufficiently large to destroy the condensate then it would have to be absorbed directly by the black hole and would perturb the geometry. These are the standard thermal channels of dissipation in which the metric perturbations $\delta g$ source temperature variations $\delta T$ of the boundary theory. Clearly in this case the whole vortex structure would be destroyed and the vortex would most likely revert to the standard case without additional core moduli. In the probe limit there is no leading order thermal dissipation since the metric is fixed to first approximation, the longitudinal modes have low energy and are absorbed by the condensate. This picture is represented in Figure 7. Therefore there should be infinitesimal changes in the vortex structure (its core shape for example) which should be visible in the dual theory. The conclusion is therefore that the dual picture of dissipation via longitudinal modes in the core is represented by small deformations in the shape of the vortex core (see Figure 8). These deformations must relax carrying energy along the plane of the vortex towards infinity. Since the system interacts very weakly with the background these excitations have long lifetimes, decaying only via thermal dissipation at higher order. Note that no dissipation is allowed through the $U(1)$ gauge-symmetry channel on the boundary since the system, breaking the dual global $U(1)$ symmetry, is in a superfluid state and hence flows without dissipation.

\section{Conclusions}

This paper investigated gravitational solutions dual to non-Abelian vortices at strong coupling and finite temperature. As usually occurs in holography, the pattern of global symmetry breaking in the dual theory is represented by gauge symmetries in the bulk. Solutions in which a neutral scalar field condenses in the core of bulk flux tubes were found and the investigation of low energy rotations of this field in internal space revealed two gapless modes. We interpreted these localised modes as the orientational moduli of the dual non-Abelian vortex. Energetic considerations revealed that this kind of solutions are preferred over the usual holographic vortices without additional moduli, proving the solution is at least metastable. Furthermore we made important connections between the bulk rotor degrees of freedom and those of the dual theory and studied some important aspects of their physics, albeit restricted by the probe limit. \newline


There are many interesting and important directions in which this work can be extended. We list below the ones which we find particularly interesting:

\begin{itemize}

\item The zero temperature $T\rightarrow 0$ limit and back-reaction: It would be ideal to be able to reach the $T =0$ limit, this necessarily involves considering the full back-reaction on the system as per \cite{Dias:2013bwa}. Having control over the full temperature range could reveal the presence of quantum phase transitions or additional finite temperature phase transitions not considered here. 

\item Finding the bulk vortex lattice. Vortex lattice solutions in holographic superconductor models are known to exist \cite{Maeda:2009vf}. Solutions such as those in this setup are important to study the coupling of the rotor degrees of freedom between vortices. Since we proved in this paper that a single vortex can confine a single rotor condensate to a spatial region it is reasonable to assume that a bulk lattice of these vortices should represent a dual lattice of rotor degrees of freedom. Then, depending on lattice spacings these degrees of freedom will interact revealing novel strong coupling physics.

\item Revealing the non-Abelian sigma model of the dual moduli: The topological considerations of the global symmetry breaking pattern reveal that the gapless orientational moduli form a non-linear $CP(1)$ sigma model. The infinitesimal parametrization for the moduli used in this paper fails to reveal how this sigma model appears in the dual theory (assuming of course that this is not spoilt in the holographic process). It would be interesting to extend this work to include the $\chi = \chi_0 S^i$ parametrization and work out the low energy bulk solutions for the $S^i$ with the constraint that $S^i S^i =1$. If a solution of this form can be found (only an analytic approach at the boundary was investigated in this paper) then this constraint should also be valid at the boundary meaning the dual moduli should also obey it. If at least at the level of vev considerations the dual fields obey this relation then it seems natural to speculate that a $CP(1)$ theory appears on the dual vortex world-line.

\item Higher dimensional extensions: this problem is related to holographic vortices in general and not simply to non-Abelian ones. It would be desirable for the study of confinement properties of these kind of solutions at strong coupling to have bulk solutions which are dual to full 3+1 dimensional flux tubes rather than 2+1 dimensional vortices. This involves adding a bulk dimension and finding a new kind of extended solitonic solution in the bulk.

\item Restricting to a single $U(1)$ gauge sector: The presence of an additional $U(1)$ gauge theory is possibly an over-complication of the system. This was recently interpreted as a dark sector in holographic vortex applications \cite{Rogatko:2015awa}. However, it seems plausible that coupling of both scalars to the original $U(1)$ gauge field might be sufficient to find non-Abelian vortex solutions. Clearly, if similar solutions to the ones found here existed with only the original $U(1)$ symmetry then the core of the vortex would also be superconducting and not correspond to a non-Abelian vortex as originally intended.

\end{itemize}

 \section*{Acknowledgments}

The author is funded by Fondecyt grant number 3140122.

\end{document}